\documentclass[aps,10pt,amsmath,amssymb,titlepage,prb,twocolumn,floatfix]{revtex4-2}

\usepackage{graphicx}% Include figure files
\usepackage{dcolumn}% Align table columns on decimal point
\usepackage{bm}% bold math
\usepackage{pgfornament}
\usepackage{braket}
\usepackage{xpatch}
\usepackage{amsmath}
\usepackage{dsfont}
\usepackage[polish,english]{babel}
\usepackage[T1]{fontenc}
\usepackage[utf8]{inputenc}
\usepackage{lmodern}
\usepackage{hyperref}

\makeatletter
\patchcmd{\@ssect@ltx}
    {\addcontentsline{toc}{#1}{\protect\numberline{}#8}}
    {}
    {}
    {}
\makeatother
\usepackage{hyperref}
\hypersetup{
    colorlinks,
    citecolor=blue,
    filecolor=black,
    linkcolor=blue,
    urlcolor=blue
}

\begin{document}
\makeatletter
\def\maketitle{
\@author@finish
\title@column\titleblock@produce
\suppressfloats[t]}
\makeatother

\preprint{APS/123-QED}

\title{Scalable entanglement of nuclear spins mediated by electron exchange}
\author{Holly G. Stemp$^{1,2}$}
    \altaffiliation[Currently at ]{Research Laboratory of Electronics, Massachusetts Institute of Technology, Cambridge, MA, USA}
\author{Mark R. van Blankenstein$^{1,2}$}
\author{Serwan Asaad$^{1,2}$}%
    \altaffiliation[Currently at ]{Currently at Quantum Machines Inc., Tel Aviv, Israel}
\author{Mateusz T. M\k{a}dzik$^{1,2}$}%
    \altaffiliation[Currently at ]{Intel Corporation Hillsboro, Oregon, United States}
\author{Benjamin Joecker $^{3}$}
\author{Hannes R. Firgau$^{1,2}$}
\author{Arne Laucht$^{1,2, 4}$}%
\author{Fay E. Hudson$^{1, 4}$}%
\author{Andrew S. Dzurak$^{1,4}$}%
\author{Kohei M. Itoh$^{5}$}%
\author{Alexander M. Jakob$^{2,6}$}%
\author{Brett C. Johnson$^{7}$}%
\author{David N. Jamieson$^{2,6}$}%
\author{Andrea Morello$^{1,2}$}%
 \email{a.morello@unsw.edu.au}

\affiliation{%
 $^{1}$ School of Electrical Engineering and Telecommunications, UNSW Sydney, Sydney, NSW 2052, Australia\\
 $^{2}$ ARC Centre of Excellence for Quantum Computation and Communication Technology\\
 $^{3}$ NNF Quantum Computing Programme, Niels Bohr Institute, University of Copenhagen, Denmark \\
 $^{4}$ Diraq Pty. Ltd., Sydney, New South Wales, Australia\\
 $^{5}$ School of Fundamental Science and Technology, Keio University, Kohoku-ku, Yokohama, Japan \\
 $^{6}$ School of Physics, University of Melbourne, Melbourne, VIC 3010, Australia\\
 $^{7}$ School of Science, RMIT University, Melbourne, VIC, 3000, Australia
}%
\date{\today}

\begin{abstract}
The use of nuclear spins for quantum computation is limited by the difficulty in creating genuine quantum entanglement between distant nuclei. Current demonstrations of nuclear entanglement in semiconductors rely upon coupling the nuclei to a common electron, which is not a scalable strategy. Here we demonstrate a two-qubit Control-Z logic operation between the nuclei of two phosphorus atoms in a silicon device, separated by up to 20 nanometers. Each atoms binds separate electrons, whose exchange interaction mediates the nuclear two-qubit gate. We prepare and measure a nuclear Bell state with a fidelity of 76  $^{+5}_{-5}\%$  and a concurrence of 0.67$^{+0.05}_{-0.05}$. With this method, future progress in scaling up semiconductor spin qubits can be extended to the development of nuclear-spin based quantum computers.

\end{abstract}

\maketitle

%\tableofcontents
Entanglement -- the most quintessential quantum mechanical property -- is a vital ingredient for quantum computation and quantum communications, and can provide a quantum advantage in sensing and metrology. However, the practical use of quantum entanglement quickly reveals a major challenge in quantum technologies, namely, the conflict between isolation from the environment and coupling between different quantum systems. In the solid state, it is generally the case that the systems with the longest coherence, such as nuclear spins \cite{saeedi2013room,zhong2015optically}, are the most challenging ones to entangle.\\

Nuclear spins were the very first physical platform to demonstrate the execution of a quantum algorithm \cite{chuang1998experimental}. Ensemble nuclear magnetic resonance (NMR) pioneered many of the sophisticated quantum control techniques that have subsequently been adopted by other platforms \cite{vandersypen2004nmr}, but is no longer pursued for scalable quantum computing, because of the lack of scalable methods to produce genuine quantum entanglement between multiple nuclei. We use the term `genuine' entanglement to highlight that, while multiple quantum coherences between distant nuclei are routinely observed in ensemble NMR \cite{warren1993generation}, such states can be described by mixtures of fully separable density matrices. A minimum spin polarization of $1/3$, i.e. a sufficiently pure initial state, is required in order to satisfy the entanglement criterion of positive partial transpose \cite{peres1996separability,simmons2011entanglement}. The development of methods to projectively measure single electron spins, and resolve nuclear spins that are coupled to them, has enabled the preparation of high-purity electron-nuclear states in e.g. diamond \cite{kolkowitz2012sensing}, silicon \cite{dehollain2016bell}, silicon carbide \cite{bourassa2020entanglement} and rare-earth systems \cite{ruskuc2022nuclear}.\\

The remaining challenge is the extremely weak mutual coupling between nuclear spins. For example, the magnetic dipole coupling between two $^{31}$P nuclei at 1~nm distance is of order 10~Hz. Therefore, most examples of nuclear entanglement rely upon coupling the nuclei to a common electron \cite{neumann2008multipartite,pfaff2013demonstration}, necessarily within distances $\sim 1-5$~nm dictated by the size of the electron wavefunction \cite{abobeih2019atomic,mkadzik2022precision,reiner2024high}. These methods are inherently non-scalable, due to exponential spectral crowding as more nuclei are included. The only way out of this gridlock is to involve multiple electrons \cite{dolde2014high,kalb2017entanglement}, which themselves should be mutually coupled in a robust and scalable manner.\\

\begin{figure*} 
    \includegraphics[width=1\textwidth]{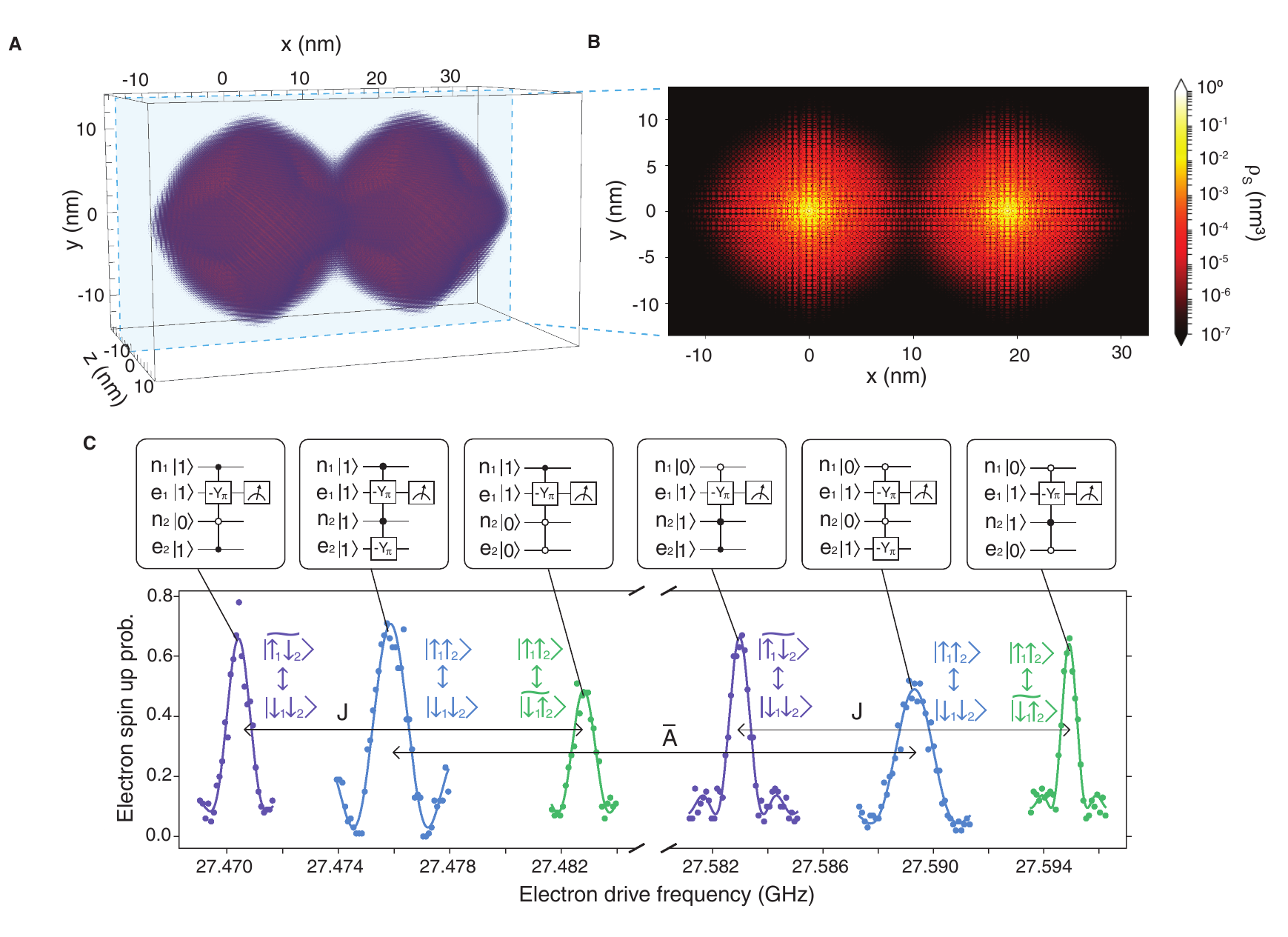}  
    \caption{\textbf{Exchange-coupled, two atom system.} 
    \textbf{A.} 3D full configuration interaction simulation of the singlet wavefunction for two donor atoms exchanged coupled with a strength of 12 MHz. This exchange coupling strength corresponds to a spacing of 19 nm along the [100] crystal axis, assuming no detuning between the two atoms. 
    \textbf{B.} 2D plane cut of the simulation in \textbf{A} for a value of z=0 nm.
    \textbf{C.} Electron spin resonance frequency spectrum schematic for one of the electrons in an exchange-coupled, two-donor system for the case of $J > \bar{A}$, where $J$ is the exchange interaction strength and $\bar{A}=(A_1 + A_2)/2$ is the average hyperfine value of the two donor atoms \cite{kalra2014robust, madzik2021conditional, stemp2024tomography}. Six resonance frequencies are present for electron 1, conditional on the state of both the other electron and the two donor nuclei. Four of these resonances represent the case for which the two donor nuclei are in an anti-parallel spin configuration, for which $J < \Delta = \bar{A}$. The remaining two of these resonances represent the case for which the two donor nuclei are in a parallel spin orientation, for which $J > \Delta = |A_{1}-A_{2}|$.}
    \label{fig:figure1} 
\end{figure*}

In this work, we demonstrate a scalable method to generate genuine entanglement between nuclear spins. We implement a nuclear two-qubit geometric controlled-Z (CZ) gate between the nuclear spins of two $^{31}$P donor atoms in a silicon metal-oxide-semiconductor (MOS) device. Each nucleus is bound to a different electron, and the two electrons are coupled by a Heisenberg exchange interaction. The exchange interaction strength of 12 MHz corresponds to an estimated inter-donor distance of up to 20 nm \cite{joecker2021full}, and enables fast ($\approx 2$~$\mu$s) entangling operations. The geometric 2-qubit gate we report here relies solely on having an electron's spin resonance frequency depending on the state of two nuclei. This condition can be fulfilled in a wide range of donor locations, and with diverse methods for electron-electron interaction, which may also involve coupling to intermediary quantum dots. Therefore, our method underpins the exciting prospect of exploiting advances in the development and scaling of semiconductor quantum dots \cite{burkard2023semiconductor,neyens2024probing}, and extending them to the operation of nuclear-spin based quantum processors.\\

\subsection*{Four-qubit, two-atom device}
The system in which these experiments were performed consists of two $^{31}$P donor nuclei, which we will henceforth refer to as donor 1 (consisting of nucleus n1 and electron e1) and donor 2 (consisting of nucleus n2 and electron e2), introduced into an isotopically purified \cite{itoh2014isotope} $^{28}$Si lattice via ion-implantation \cite{jakob2022deterministic}. We use the symbols $\ket{\downarrow},\ket{\uparrow}$ for the electron spin states, and $\ket{\Downarrow},\ket{\Uparrow}$ for the nuclear ones. Each donor nucleus is hyperfine coupled to a single bound electron, with a hyperfine coupling strength of $A_{1}=$ 111 MHz and $A_{2}=$ 113 MHz, respectively. The two electrons are coupled to one another with an exchange interaction strength of $J\approx12$~MHz \cite{madzik2021conditional}. Using a full configuration interaction method \cite{joecker2021full} to model the two donor atoms (as shown in Fig. \ref{fig:figure1} \textbf{A,B}) we estimate the largest inter-donor distance for this value of exchange interaction strength to be 20 nm. The estimated inter-donor distance for this value of exchange interaction strength depends upon the exact axis over which the donors are separated. Averaging over all possible donor orientations we obtain an average estimated inter-donor distance of 16.13 nm. For details of this calculation see section `Inter-donor distance estimations' in the Supplementary Material. These calculations assume no electrical detuning between the donors, which would increase the inter-donor distances possible for this exchange interaction strength. \\

The spin Hamiltonian of the two-donor electron-nuclear system, in frequency units, is:

\begin{align}
        H = &(\mu_{\text{B}}/h) B_{0}(g_{1}S_{z1} + g_{2}S_{z2})+\\
    & \gamma_\mathrm{n}B_{0}(I_{z1}+I_{z2})+ \nonumber \\
    &  A_{1}\mathbf{S_{1}}\cdot \mathbf{I_{1}} +  A_{2}\mathbf{S_{2}}\cdot \mathbf{I_{2}} + \nonumber \\
    & J(\mathbf{S_{1}\cdot S_{2}}), \nonumber
\end{align}
where $\mu_{\text{B}}$ is the Bohr magneton, $h$ is Planck's constant, $g_{1,2}\approx 1.9985$ the Land\'e  g-factors of each electron spin, $g\mu_{\text{B}}/h \approx$ 27.97 GHz/T and $\gamma_\mathrm{n} \approx$ 17.23 MHz/T is the $^{31}$P nuclear gyromagnetic ratio. For information on how the donor nuclei and electrons are initialized, controlled and read out see the `Methods and Materials' section of the Supplementary Material \cite{methods}. For coherence time measurements of the ionized and neutral nucleus and the donor electron, see section `Nuclear and electron coherence times' in the Supplementary Material. \\

Within each donor, the hyperfine coupling results in two resonance frequencies being present for each spin, conditional on the other spin being in either the $\Downarrow$($\downarrow$) or $\Uparrow$($\uparrow$) state. Therefore, any operation on either the electron or nucleus represents a two-qubit conditional rotation (CROT) gate, which can be used to generate electron-nuclear entanglement \cite{dehollain2016bell}. \\

This entanglement can be extended to multiple donor atoms by introducing an exchange coupling between the electrons of separate donor atoms. Figure \ref{fig:figure1} \textbf{C} shows the ESR frequency spectrum for one of the electrons in the two donor system in the presence of an exchange interaction between the electrons. The insets in the ESR spectrum show the electron state transitions represented by each resonance frequency peak. The presence of the exchange interaction causes the electron eigenstates to hybridize, such that the two-qubit electron eigenstates become: $\ket{\downarrow_1 \downarrow_2}, \widetilde{\ket{\downarrow_1 \uparrow_2}} = \cos{(\theta)}\ket{\downarrow_1 \uparrow_2} + \sin{(\theta)}\ket{\uparrow_1 \downarrow_2}, \widetilde{\ket{\uparrow_1 \downarrow_2}} = \cos{(\theta)}\ket{\uparrow_1 \downarrow_2} - \sin{(\theta)}\ket{\downarrow_1 \uparrow_2}, \ket{\uparrow_1 \uparrow_2}$, where $\tan(2\theta) = \frac{J}{\Delta}$ and $\Delta$ is the detuning between the two electrons dependent on the orientation of the nuclear spins. In this case, each electron resonance depends on the state of three spins: the two donor nuclei and the second electron of the exchange-coupled pair. Each rotation therefore constitutes a 4-qubit Toffoli gate. This gate has two main advantages. First, it allows us to generate entanglement between a donor electron and the nucleus of another donor atom. Secondly, this gate allows us to implement a two-qubit nuclear geometric controlled-Z (CZ) gate; thus enabling entanglement between neighbouring donor nuclei.\\

\begin{figure*} 
    \centering
    \includegraphics[width=1\textwidth]{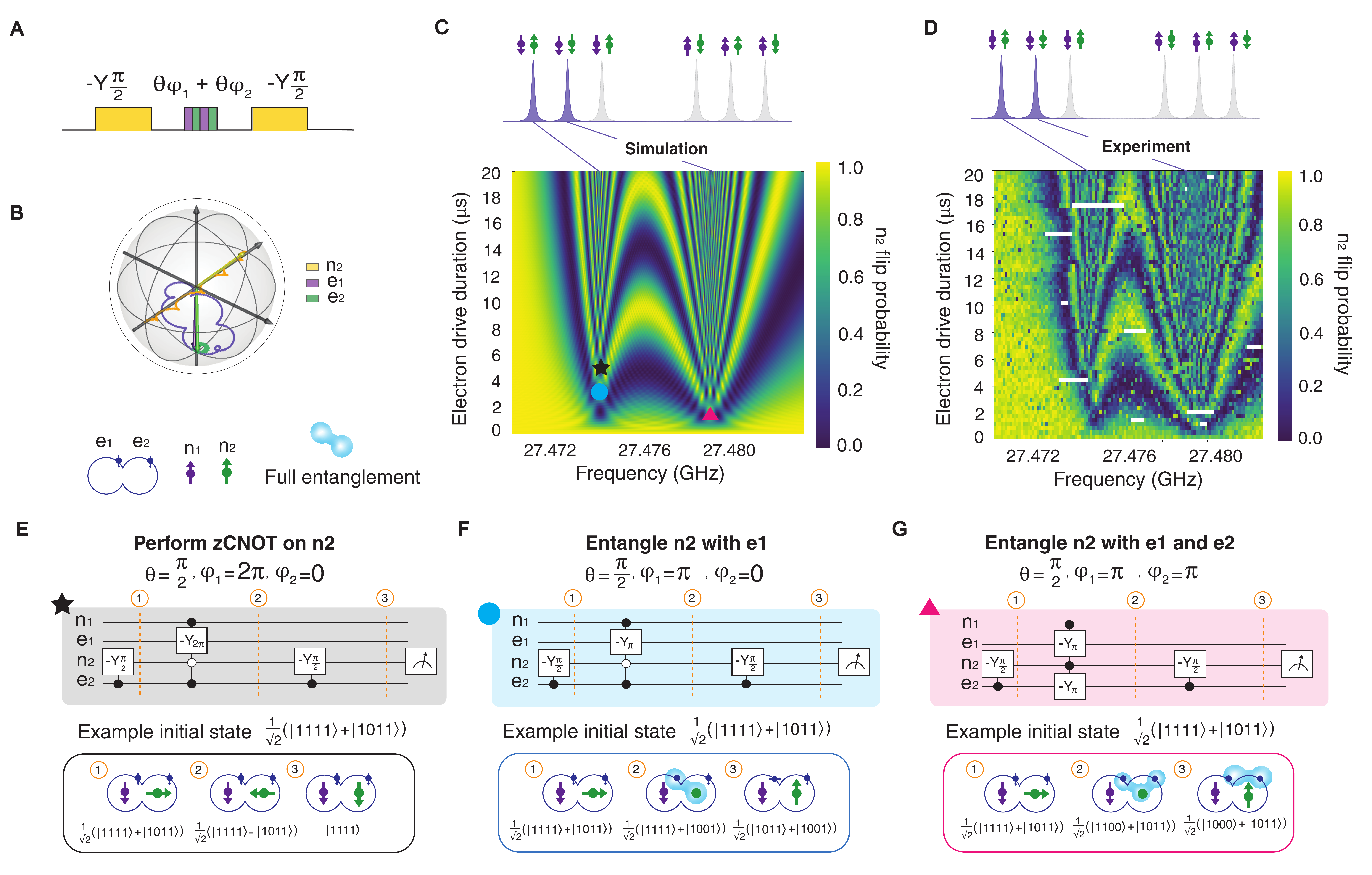} 
    \caption{\textbf{Entangling electrons and nuclei.} 
    \textbf{A.} Schematic of the pulse sequence used to generate the plots in \textbf{C} and \textbf{D}. The yellow pulses represent a $-Y_{\frac{\pi}{2}}$ performed on n2. The purple and green striped pulse represents a pulse of varying duration and frequency performed on the electrons in the J-coupled system.  
    \textbf{B.} Simulation of the path of n2 (yellow), e1 (purple) and e2 (green) on the Bloch sphere for the case in which the electron drive is conditional on the nuclei being in the $\ket{\Downarrow_1 \Uparrow_2}$ state. Here the arrows mark the final state of the spins following the drive. For a drive duration of $\pi$ the arrows of n2 and e1 shrink to the centre of the sphere, as the two spins become entangled. 
    \textbf{C.} Simulation of the pulse sequence in \textbf{A} where the axes  correspond to the frequency and duration of the pulse applied to the electrons. 
    \textbf{D.} Experimental data from the implementation of the pulse sequence in \textbf{A}. 
    \textbf{E-G.} Circuit diagrams and example input and output states representing the operations performed at the points in \textbf{C} with \textbf{E}, the black star, \textbf{F}, the blue circle, and \textbf{G}, the pink triangle. Each state is written in the format $\ket{\text{n}_{1}\text{n}_{2}\text{e}_{1}\text{e}_{2}}$. Black dots in the circuit diagram represent operations conditional on the $\ket{1} = \ket{\downarrow}$ state, while white dot represents operations conditional on the state $\ket{0} = \ket{\uparrow}$. } 
    \label{fig:figure2} 
\end{figure*}

\subsection*{The nuclear geometric CZ gate in a J-coupled system}\label{s:CZ_gate}
The only requirement for being able to implement a nuclear geometric CZ gate is the presence of a resonance frequency of an electron, which depends on the state of two nuclei. This can be achieved in this system using the 4-qubit Toffoli gate discussed above, with the following steps:\\

First, both nuclei are initialized in the spin $\ket{\Downarrow}$ state using the ENDOR scheme described in \cite{stemp2024tomography}. Next, one of the two nuclei is placed in a superposition state via an NMR $-Y_{\frac{\pi}{2}}$ pulse, conditional on the electron bound to that nucleus being in the $\ket{\downarrow}$ state. For the example of n2 being placed in a superposition state, this therefore results in the following nuclear state \cite{mkadzik2022precision}

\begin{equation}
    \label{nuclear_state}
    \ket{\Downarrow_1} \otimes \frac{1}{\sqrt{2}}(\ket{\Downarrow_2} + \ket{\Uparrow_2}) = \frac{1}{\sqrt{2}} (\ket{\Downarrow_1 \Downarrow_2} + \ket{\Downarrow_1 \Uparrow_2}).
\end{equation}

When a 2$\pi$ rotation is then applied to one of the electrons, conditional on a given two-qubit nuclear state, this rotation imparts a geometric phase, $\phi_{\rm{G}}$ onto the two-qubit nuclear state upon which the electron rotation was conditioned, with the magnitude of the geometric phase imparted given by

\begin{align}
    \phi_{\rm{G}} = -\frac{1}{2}\Delta \Omega,
\end{align}

where $\Delta \Omega$ represents the angle subtended on the sphere by the trajectory of the electron spin \cite{nietot1992aharonov}. For the case of $\Delta \Omega = 2\pi$, $\phi_{\rm{G}} = -\pi$ and hence a phase of $\pi$ is acquired by the two-qubit nuclear state upon which the electron rotation was conditioned, relative to the other states in the nuclear superposition. For the case of the electron rotation being conditioned on the two-qubit nuclear state $\ket{\Downarrow_1 \Uparrow_2}$, this would therefore result in the following

\begin{equation}
    \label{nuclear_state_2}
    \frac{1}{\sqrt{2}} (\ket{\Downarrow_1 \Downarrow_2} + \ket{\Downarrow_1 \Uparrow_2}) \xrightarrow[]{\phi_{\rm{G}}}\frac{1}{\sqrt{2}} (\ket{\Downarrow_1 \Downarrow_2} - \ket{\Downarrow_1 \Uparrow_2}).
\end{equation}

Note that if n1 was instead initialized in the $\ket{\Uparrow_{1}}$ state, then the $2\pi$ rotation of the electron would be far off-resonance and hence no geometric phase would be imparted. This gate therefore represents the implementation of a two-qubit nuclear CZ gate, where a phase of $\pi$ is added to the target nucleus, conditional on the control nucleus being in the $\ket{1} = \ket{\Downarrow}$ state. \\

A controlled-rotation (CNOT) or zero-controlled-rotation (zCNOT) nuclear gate can be implemented by adding an extra $\frac{\pi}{2}$ pulse to the target nucleus after the CZ gate.\\

\subsection*{Entangling electrons and nuclei}
By varying the frequency and duration of the drive applied to the electrons during the implementation of the nuclear geometric CZ gate, entanglement can be generated between one of the nuclei and both electrons in the exchange-coupled pair. Fig. \ref{fig:figure2} shows the simulated (\textbf{C}) and experimental (\textbf{D}) results of the implementation of the pulse sequence shown in Fig. \ref{fig:figure2} \textbf{A}. Before the beginning of the sequence, the two nuclei are initialized in the state $\ket{\Downarrow_1 \Downarrow_2}$. A ${\rm -Y}_{\frac{\pi}{2}}$ pulse is applied to n2, in order to prepare the nuclear superposition state $\frac{1}{\sqrt{2}}(\ket{\Downarrow_1 \Downarrow_2} + \ket{\Downarrow_1 \Uparrow_2})$. A pulse of varying duration and frequency is then applied to the electrons. The frequency of this ESR pulse is swept across a range of 20 MHz, such that it crosses the resonance frequencies of e1 conditional on the $\ket{\Downarrow_1 \Uparrow_2}$ nuclear state and of e1 and e2 conditional on the $\ket{\Downarrow_1 \Downarrow_2}$ nuclear state, which together make up the nuclear superposition. Similarly, the duration of the ESR pulse is swept from 0 $\mu$s to approximately 20 $\mu$s, or approximately 20 $\pi$ pulses of the electron when on resonance. A final $\frac{\pi}{2}$ pulse is then applied to n2. This acts to project the information along the XY plane of the Bloch sphere to the Z-axis, such that it can be read out. For this measurement we extracted the nuclear flipping probability, $P_{\rm flip}$. Fig. \ref{fig:figure2} \textbf{B} shows the simulated trajectory on the Bloch sphere of n2 and e1, for the case of a $2\pi$ rotation of the electron performed on resonance. \\

In Fig. \ref{fig:figure2} \textbf{E-G}, we highlight three points of note in the results of Fig. \ref{fig:figure2} \textbf{C}. Fig. \ref{fig:figure2} \textbf{E} highlights a point at which e1 undergoes a 2$\pi$ rotation at its resonance frequency (black star in Fig. \ref{fig:figure2} \textbf{C}). This rotation is conditional on one of the states of the nuclear superposition: $\ket{\Downarrow_1 \Uparrow_2}$. When the nuclei are in an anti-parallel spin orientation, the detuning between e1 and e2 (given by $\Delta = \bar{A}  = 112$ MHz, where $\bar{A}$ is the average hyperfine value of the donor atoms) ensures that only e1 is driven at this frequency. As indicated by the circuit diagram, this 2$\pi$ electron rotation therefore corresponds to the implementation of the nuclear geometric CZ gate, as described in the previous section. Since the operation is physically applied to the electron, it proceeds at the speed of electron rotations, much faster than nuclear rotations \cite{morton2006bang}. In this case, with a moderate power of 15 dBm at the output of the microwave source, the nuclear CZ gate is completed in $\approx 2$~$\mu$s. When followed by the final $\frac{\pi}{2}$ pulse on n2, this operation performs a zCNOT gate, flipping n2 if n1 is in the $\ket{\Uparrow} = \ket{0}$ state.\\

Figure \ref{fig:figure2} \textbf{F}, shows the second point of note in Fig. \ref{fig:figure2} \textbf{C}. This point represents the implementation of a $\pi$ pulse on e1, once again at the ESR resonance frequency conditional on the nuclear state $\ket{\Downarrow_1 \Uparrow_2}$ (blue circle in Fig. \ref{fig:figure2} \textbf{C}). In this case, a $\pi$ rotation of the electron will flip the state of e1, conditional on the state of n2, which is in a superposition state. This operation therefore acts to maximally entangle e1 with n2. \\

Finally, Fig. \ref{fig:figure2} \textbf{G}, highlights the point for which both electrons undergo a $\pi$ rotation, this time conditional on the nuclear state $\ket{\Downarrow_1 \Downarrow_2}$ (pink triangle in Fig. \ref{fig:figure2} \textbf{C}), which constitutes the other half of the prepared nuclear superposition state. For the case of the nuclei being in a parallel spin orientation, $J > \Delta$, where $\Delta$ is now given by the detuning between the two electrons $\Delta = |A_1 - A_2| = 2$ MHz. In this case the eigenstates of the two electrons become strongly hybridized with one another \cite{kalra2014robust, madzik2021conditional, stemp2024tomography}, and hence driving at this frequency results in both e1 and e2 undergoing a rotation of $\pi$ conditioned on n2. This results in both electrons becoming maximally entangled with n2, thus generating a 3-qubit entangled state. \\

The phase imparted on n2 by the electrons depends on a combination of both the frequency detuning ($\Delta_f = f_{0} - f_{\rm MW}$ where $f_0$ is the resonance frequency and $f_{\rm MW}$ is the frequency of the driving field) and the duration of the electron drive. As $\Delta_f$ increases, the area of the cone enclosed by the path of the electron traversed on the Bloch sphere decreases. However, as the detuning from the resonance frequency increases, the spin precession frequency in the rotating frame increases, resulting in a greater number of rotations about this cone. There are therefore a number of regions in Fig. \ref{fig:figure2} \textbf{C} and \textbf{D} away from the electron resonance frequencies, for which the electron imparts a phase of $2\pi$ on the nucleus or undergoes a rotation of $\pi$, resulting in either the implementation of a nuclear CZ gate, or the generation of entanglement between the electrons and n2. A full simulation of this pulse sequence is shown in Supplementary Materials section `Phase map simulations'.\\

\begin{figure*} % Do NOT use \begin{figure*}
    \centering
    \includegraphics[width=1\textwidth]{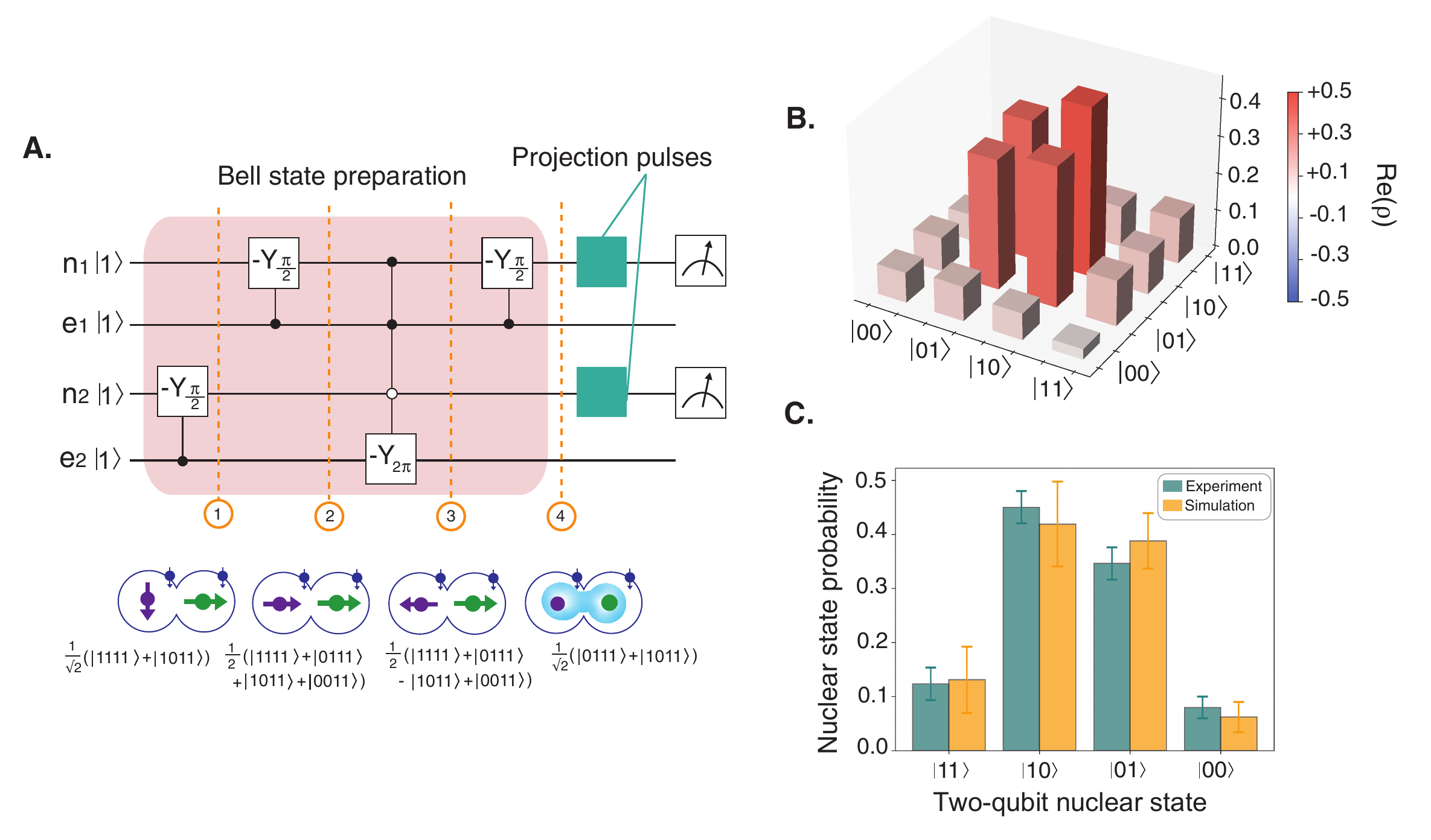}  
    \caption{\textbf{Long-range entanglement between two nuclei.} 
    \textbf{A.} Circuit diagram depicting the pulse sequence used to generate the Bell state $\frac{1}{\sqrt{2}}(\ket{01} + \ket{10})$ between the two nuclei (red-shaded region). A zCNOT is performed on n1 by sandwiching the nuclear geometric CZ gate between two $\frac{\pi}{2}$ pulses. The pulses depicted in the turquoise box above the circuit diagram show the projection pulses performed after the Bell state generation, such that the Bell state can be measured along all axes of the Bloch sphere, which is a requirement to perform Bell state tomography. 
    \textbf{B.} Reconstructed density matrix of the nuclear Bell state constructed using Bell state tomography. This density matrix was reconstructed without removing SPAM errors. 
    \textbf{C.} Experimental and simulated results of the two-qubit measurement of the Bell state along the ZZ axis. The simulation assumed perfection in the gate operations and readout, with the only error introduced being the electron initialization error measured from experiments.}
    \label{fig:figure3} 
\end{figure*}

\subsection*{Long-range nuclear entanglement}
Using the nuclear geometric CZ gate, we prepare an entangled state between the two distantly spaced donor nuclei. Fig. \ref{fig:figure3} \textbf{A}, shows the pulse sequence used to prepare a Bell state between the two nuclei. The nuclei are first initialized in the state $\ket{\Downarrow_1 \Downarrow_2}$. A $\frac{\pi}{2}$ NMR pulse is then applied to both of the nuclei sequentially, such that we prepare the nuclear superposition state $\frac{1}{2}(\ket{\Downarrow_1 \Downarrow_2} + \ket{\Downarrow_1 \Uparrow_2} + \ket{\Uparrow_1 \Downarrow_2} + \ket{\Uparrow_1 \Uparrow_2})$. A $2\pi$ ESR pulse is then applied to e2 conditional on the nuclear state $\ket{\Downarrow_1 \Uparrow_2} = \ket{10}$, which imparts a phase of $-\pi$ to this component of the nuclear superposition, resulting in the state $\frac{1}{2}(\ket{\Downarrow_1 \Downarrow_2} - \ket{\Downarrow_1 \Uparrow_2} + \ket{\Uparrow_1 \Downarrow_2} + \ket{\Uparrow_1 \Uparrow_2})$. A second $\frac{\pi}{2}$ pulse is then applied to n1, yielding the final state $\frac{1}{\sqrt{2}}(\ket{\Uparrow_1 \Downarrow_2} + \ket{\Downarrow_1 \Uparrow_2}) = \frac{1}{\sqrt{2}}(\ket{01} + \ket{10})$, which represents the $\Psi^{+}$ Bell state between the two nuclei. The resulting nuclear state probability is then measured for both nuclei. \\

In order to quantify the degree of entanglement generated between the donor nuclei we performed nuclear Bell state tomography. This involved first preparing the nuclear Bell state using the pulse sequence detailed above, before projecting the Bell state information along different axes of the Bloch sphere to the Z-axis for read out \cite{james2001measurement}. The pulse sequence used to prepare the nuclear Bell state is shown in Fig. \ref{fig:figure3} \textbf{A}. This information was then used to reconstruct the density matrix of the prepared Bell state (see Supplementary Materials section `Bell state fidelity and concurrence'), shown in Figure \ref{fig:figure3} \textbf{B}. From this density matrix, a Bell state fidelity of 76  $^{+5}_{-5}\%$ and concurrence value of 0.67 $^{+ 0.05}_{- 0.05}$ were obtained, without removal of SPAM, indicating the clear presence of entanglement between the two nuclei. The error bars here and throughout this work represent $2\sigma$. For more information on how the error bars were calculated, see Supplementary Materials section `Error bar estimation'. \\

The nuclear Bell state fidelity is primarily limited by the fidelity of electron initialization. The nucleus can be read out with high fidelity ($> 99\%$ \cite{pla2013high}) due to the ability to perform QND readout via the electron. Additionally, the operation fidelity of the nucleus, measured using single-qubit gate set tomography (see Supplementary Materials section `Nuclear one-qubit gate set tomography (GST)'), also exceeds 99$\%$. Consequently, the reduction in the amplitude of a Rabi oscillation performed on the neutral nucleus is primarily caused by electron initialization errors. By analyzing the amplitude of the neutral nucleus Rabi oscillations, we estimate an electron initialization fidelity of 86$\%$ (see Supplementary Materials section `Phase reversal tomography'). This initialization error significantly impacts the performance of initializing and operating the neutral nuclei, as each nuclear operation is conditional on the electron state.\\

Figure \ref{fig:figure3} \textbf{C} shows the experimentally measured two-qubit state probabiliy of the prepared nuclear Bell state along the Z-axis of the Bloch sphere, without any projection pulses applied. This plot includes simulated two-qubit state probability, assuming perfect gates and readout, with the only introduced error being the experimentally informed electron initialization error. The overlap between the measured data and simulation results highlights the significant role of electron initialization in limiting Bell state fidelity. The root cause of this problem is poor thermalization of the electron reservoir from which the $\ket{\downarrow}$ state is initialized \cite{johnson2022beating}. It is usually possible to solve this by using a Bayesian Maxwell's demon to beat the thermal limit of electron initialization~\cite{johnson2022beating}. However, in this particular device, the method did not result in any improvements, due to the short electron tunnelling times (see also Supplementary Materials section `Electron initialization'). \\

Previous implementations of Maxwell's demon on donor devices with longer electron tunneling times have demonstrated an electron initialization fidelity of 98.9$\%$. With improvements to the amplifier chain to increase readout bandwidth, this fidelity could be realistically increased to 99.9$\%$ \cite{johnson2022beating}. Assuming this electron initialization fidelity, we obtain a simulated nuclear Bell state fidelity, using the same simulation as described above, of 99.7$\%$. Using deterministic single-ion implantation \cite{jakob2022deterministic} to better control the distance between donor and SET could further improve this value.

\subsection*{Conclusion}
Entanglement is a vital resource for quantum computing. This work represents a key milestone, demonstrating the entanglement of nuclear spin qubits with a scalable method. Although different in the details of the operation, this result also represents the first realization of the original vision of Kane's silicon-based quantum computer \cite{kane1998silicon}, where electron exchange was the key ingredient to provide universal quantum logic between nuclear spin qubits.\\

In the future, the distance between entangled nuclei could be further extended by adopting other methods of coupling the electron spins. This can be achieved for example by using a large jellybean quantum dot positioned between the donors to mediate the exchange interaction \cite{wang2023jellybean}, or by using a superconducting resonator to mediate this coupling \cite{xu2023coupling}. These increased length scales could increase the viability of donor spins in silicon as a scalable quantum processor architecture. Promising experiments have already shown the ability to create large, deterministic donor arrays on a 300~nm pitch by ion implantation \cite{jakob2024scalable}. The fact that our results were obtained using a MOS-compatible silicon device, where ion-implanted donors are integrated within the same device structure adopted in lithographic MOS dots \cite{veldhorst2015two}, will allow us to link progress in nuclear-spin based quantum computing to the burgeoning field of semiconductor quantum dots \cite{burkard2023semiconductor}.

\section*{Acknowledgments}
\paragraph*{Funding:}
This research was funded by the Australian Research Council Centre of Excellence for Quantum Computation and Communication Technology (CE170100012) and the US Army Research Office (Contracts no. W911NF-17-1-0200 and W911NF-23-1-0113). A.M. acknowledges an Australian Research Council Laureate Fellowship (FL240100181). We acknowledge the facilities, and the scientific and technical assistance provided by the UNSW node of the Australian National Fabrication Facility (ANFF), and the Heavy Ion Accelerators (HIA) nodes at the University of Melbourne and the Australian National University. ANFF and HIA are supported by the Australian Government through the National Collaborative Research Infrastructure Strategy (NCRIS) program. H.G.S., M.R.v.B. acknowledge support from the Sydney Quantum Academy. The views and conclusions contained in this document are those of the authors and should not be interpreted as representing the official policies, either expressed or implied, of the Army Research Office or the U.S. Government. The U.S. Government is authorized to reproduce and distribute reprints for Government purposes notwithstanding any copyright notation herein.
\paragraph*{Author contributions:}
H.G.S., M.R.vB., S.A., and A.M. conceived and designed the experiments. H.G.S., M.R.vB. and H.R.F. performed and analysed the measurements. M.T.M. and F.E.H. fabricated the device, with A.S.D.'s supervision, on materials supplied by K.M.I.. A.M.J., B.C.J. and D.N.J. designed and performed the ion implantation. B.J. performed full configuration interaction simulations. A.L. assisted with Floquet time evolution simulations. H.G.S., M.R.vB. and A.M. wrote the manuscript, with input from all coauthors. A.M. supervised the project.
\paragraph*{Competing interests:}
A.S.D. is the CEO and a director of Diraq Pty Ltd. A.L., F.E.H. and A.S.D. declare equity interest in Diraq Pty Ltd.
\paragraph*{Data and materials availability:}
All data are available in the manuscript, the supplementary material or deposited at Dryad \cite{stemp2025scalabledata}.

% \bibliography{apssamp}


\twocolumngrid
\newpage
\clearpage

\onecolumngrid
\title{Supplementary Information: Scalable entanglement of nuclear spins mediated by electron exchange}

\maketitle
\renewcommand{\thepage}{S\arabic{page}}
\renewcommand{\thetable}{S\arabic{table}}
\renewcommand{\thefigure}{S\arabic{figure}}
\renewcommand{\theequation}
{S\arabic{equation}}
\setcounter{figure}{0}
\setcounter{section}{0}
\onecolumngrid

\begin{center}
    \large{CONTENTS}
\end{center}

\renewcommand{\arraystretch}{2}
\begin{tabular}{m{30em}m{17em}m{1.5em}}
Materials and Methods & & S11\\
Pulse-induced resonance shift (PIRS) &  & S12 \\
Flip probability vs up proportion &  & S14 \\
Nuclear one-qubit gate set tomography (GST) &  & S15 \\
Phase map simulation &  & S15 \\
Bell state fidelity and concurrence &  & S16 \\
Error bar estimation & & S20 \\
Phase reversal tomography &  & S20 \\
Electron initialization &  & S24 \\
Nuclear and electron coherence times &  & S25 \\
Inter-donor distance estimations &  & S25 \\
Compatibility of this nuclear coupling scheme with deterministic ion implantation &  &  S26
\end{tabular}

\clearpage

\subsection*{Materials and Methods}
\subsubsection*{Fabrication}
For the device fabrication, an isotopically enriched epilayer of $^{28}$Si, with a residual $^{29}$Si concentration of 730 ppm, is grown atop a silicon wafer. A 200 nm thick SiO$_{2}$ field oxide is grown via a wet oxidation process, while in the active region of the device a high-quality 8 nm thermal oxide layer is grown in dry conditions. In order to implant the devices with donor atoms, a 90 nm x 100 nm implantation window is defined using electron-beam lithography (EBL). Donor atoms are then implanted into this window at an acceleration voltage of 10 kV and implantation dose of  1.4 $\times$ 10$^{12}$ cm$^{-2}$. Using SRIM simulations we have estimated that at this implantation energy, the implantion profile of the donors should exist in a range approximately 10 nm  from the oxide interface. A rapid thermal anneal for 5 seconds at 1000 $^{o}$C is carried out following implantation in order to repair the damage inflicted to the lattice during implantation and to activate the donors. Aluminium gates are then fabricated on the surface of the chip in three separate EBL steps, with each step being followed by a thermal deposition of aluminium, followed by exposure to pure, low pressure oxygen to form an Al$_{2}$O$_{3}$ layer between each gate layer. A forming gas anneal (95$\%$ N$_{2}$, 5$\%$ H$_{2}$) for 15 minutes at 400 $^{o}$C is then performed in order to passivate any interface traps. 
\subsubsection*{Experimental setup}
The device is glued to a copper enclosure, through a cut out in a custom printed circuit board (PCB), that contains both microwave and low frequency lines. The device is then aluminium wire bonded to the PCB and bolted to a permanent magnet assembly, which provides a strong, static magnetic field. The permanent magnet assembly is mounted onto the mixing chamber plate of a Bluefors BF-LD400 cryogen free dilution refrigerator, which cools the device to a base temperature of approximately 20 mK.\\

DC voltages are sourced by 9 Stanford Research systems (SRS) SIM 928 DC sources, hosted in two SIM 900 mainframes. The DC lines consist of a constantan loom, which is low pass filtered with a 20 Hz cutoff at the mixing chamber stage. Three of the donor gates also have an AC input to allow for fast dynamic tuning of these gate voltages. These RF signals are provided by a Keysight M3300A arbitrary waveform generator (AWG), which is bandwidth limited to 200 MHz, and passed into the AC inputs of AC/DC combiners for these gates, to allow an RF modulation to be added on top of the static DC voltage. The RF lines are made from a flexible copper-nickel (Cu-Ni) coaxial line. These lines are graphite coated to reduce triboelectric noise \cite{kalra2016vibration} and filtered to a cutoff frequency of 145 MHz at the mixing chamber plate.\\

A Keysight M3201A module hosts a field-programmable gate array (FPGA), on which we
have built an in-house direct digital synthesis (DDS) system. The DDS provides RF input signals to the I and Q inputs of a Keysight E8267D PSG vector microwave source, to allow for single and dual-sideband modulation of a microwave tone produced by the vector source, for control of the electron spin. Additionally, the DDS provides the RF signal for control of the nuclear spin through NMR. The NMR signal is attenuated by 10 dB at room temperature, to protect the antenna from any unexpected spikes in NMR amplitude, before being combined with the microwave output of the vector source using a DPX1721 diplexer, allowing both NMR and ESR signals to travel down the same line to the device’s antenna. This line consists of a semi-rigid, silver-plated Cu-Ni coaxial line. The line then passes through an inner/outer DC block at room temperature, to prevent possible voltage offsets from creating a DC current through the thin short circuit termination of the microwave antenna. Additionally, this line is attenuated by a further 10 dB, with an attenuator positioned at the 4K stage of the dilution refrigerator.\\

The SET current returning from the drain is passed through a Basel SP983c transimpedence amplifier, which converts the current into a voltage with a gain of 10$^{7}$ V/A and frequency bandwidth of 100 kHz. The voltage signal is then passed into a SIM911 bipolar junction transistor (BJT) amplifier, which is housed in the same SIM900 mainframe as the DC sources and provides an additional gain of 10$^{2}$. This amplifier also has the function of breaking the ground between the fridge and the measurement setup. As the BJT amplifier can add additional noise to the current signal, the signal is passed through a further passive low pass filter, with a cutoff at 100 kHz. The signal then enters the digitiser channel of the Keysight M3300A, with a sample rate of 100 megasamples-per-second.\\

Note that for the phase reversal tomography experiments in the Supplementary Material, a Quantum Machines OPX+ was used as an AWG and digitiser, instead of the Keysight M3201A and Keysight M3300A. The OPX+ has a sample rate of 2 gigasamples-per-second.

\subsubsection*{Donor readout and initialization}
The electron spin of donor 1 is read out and initialized via spin-dependent tunneling to a nearby single-electron transistor (SET) island \cite{morello2010single}. The electron spin of donor 2 on the other hand, is read out indirectly via electron 1, as described in detail in Ref. \cite{stemp2024tomography}. The two donor nuclei, n1 and n2, are also read out indirectly via e1. This process involves rotating e1 conditional on the state of the hyperfine-coupled nucleus, before reading the electron’s state. Although in this experiment we perform four sequential pulses and readouts of the electron, conditional on the four two-qubit nuclear states, we also have the option to perform parity readout of the nuclei in this system. This can be done by applying two pulses on the electron simultaneously, conditional on, for example, the nuclear states $\ket{\Downarrow_1 \Uparrow_2}$ and $\ket{\Uparrow_1 \Downarrow_2}$ followed by an electron readout. This operation constitutes a single-step parity check (SSPC) gate, a useful operation for quantum error correction \cite{ustun2024single}. We perform multiple quantum non-demolition (QND) readouts on the nuclei \cite{pla2013high,joecker2024error} to reduce the overall readout error. Each repetition of performing a given pulse sequence, followed by a nuclear QND readout is referred to as a ‘shot’. Throughout this work we report either nuclear spin up proportion, $P_{\Uparrow}$, (or state probability for 2-qubit readout) or nuclear flip probability, $P_{\rm flip}$, i.e. the probability that the nuclear state flips from one shot to the next. For further details regarding the nuclear spin measurements, refer to Supplementary Materials section `Flip probability vs up proportion'. The nuclei are initialized via a well-established electron-nuclear double resonance (ENDOR) technique, consisting of a series of pulses on both the electrons and nuclei which leave the nucleus in the desired state regardless of its initial state \cite{tyryshkin2006davies}.\\

The spin of the electrons and nuclei are coherently manipulated using electron spin resonance (ESR) or nuclear magnetic resonance (NMR), respectively \cite{pla2012single, pla2013high}. This is achieved by applying oscillating magnetic fields of either microwave or radio frequency (RF) via an on-chip broadband antenna.

%%%%%%%%%%%%%%%% SUPPLEMENTARY TEXT %%%%%%%%%%%%%%%
\subsection*{Pulse-induced resonance shift (PIRS)}

\begin{figure}[ht]
    \centering
    \includegraphics[width=0.7\textwidth] {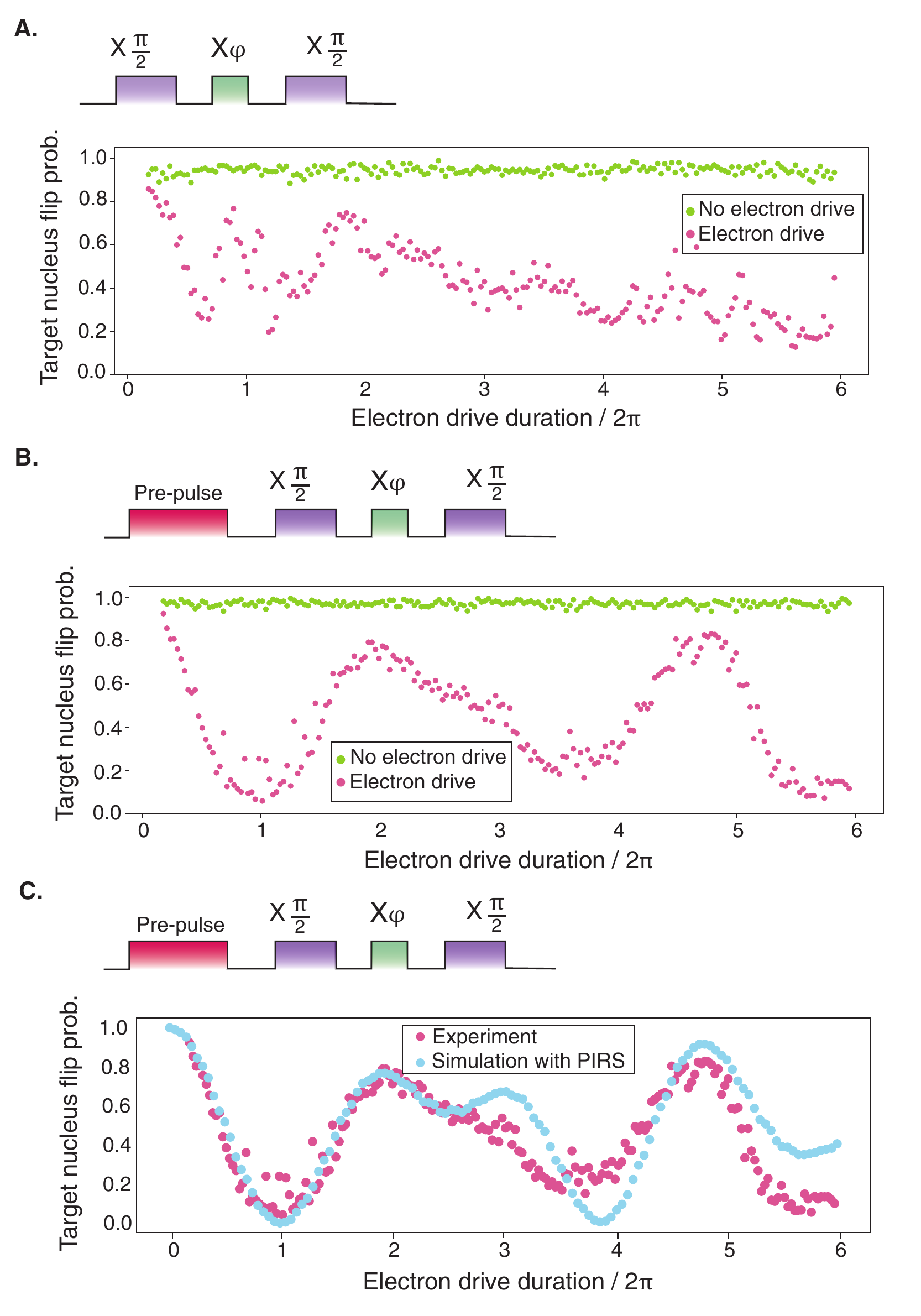}
    \caption[Pulse induced resonance shift (PIRS)]{\textbf{Pulse induced resonance shift (PIRS) and the geometric CZ gate.} \textbf{A.} Experimental results of the pulse sequence shown above the plot either including (pink points) or excluding (green points) the rotation of the electron. \textbf{B.} The implementation of the sequence in \textbf{A.} but with a saturation pulse applied before the start of the sequence in this case, which consists of a long ($\approx 5$ times the $\pi$ time of the nucleus) NMR pulse which is not in resonance with any NMR transition frequencies. \textbf{C.} Experimental (pink points) and simulated (blue points) results of the pulse sequence shown above the plot with PIRS.}
    \label{fig:pulse_induced}
\end{figure}

The fidelity of the nuclear geometric CZ gate is highly sensitive to the microwave pulse driving the electron rotation, which must be precisely tuned to the resonance frequency. This is because achieving an exact $\pi$ geometric phase on the nucleus requires the electron to complete an exact $2\pi$ rotation during the ESR pulse.\\

A key source of fidelity loss is the observed shift in the electron's resonance frequency upon applying an NMR pulse, a phenomenon known as pulse-induced resonance shift (PIRS) \cite{freer2017single}. There are a number of notable characteristics of PIRS that clearly differentiate this effect from other causes of pulse-induced frequency shifts, such as the AC Zeeman shift. These characteristics include:

\begin{enumerate}
    \item The electron resonance frequency does not shift instantaneously upon switching on the AC driving field but instead drifts over a timescale of $\approx 100 \mu$s, before saturating at a new frequency approximately $100$ kHz from the original resonance frequency. Similarly, upon switching off the AC driving field, the electron resonance drifts back to its original resonance frequency over the same timescale.
    \item The shift in the electron resonance frequency does not depend on the frequency of the AC driving field  but rather the power and duration for which it is applied. This results in a shift in the resonance frequency of the electrons occurring when an NMR pulse is applied, despite the NMR pulse frequency being detuned from the ESR resonance by $\approx 27$ GHz. The reason this effect is more prevalent during an NMR pulse compared to an ESR pulse, is due to the fact that the NMR pulse is typically applied at $\approx$ 10 times the amplitude of the ESR pulse and for longer durations.
\end{enumerate}

PIRS is not an effect unique to this device, nor even to the donor system more generally; the effect has been reported in a number of different semiconductor qubit devices with a variety of architectures and material platforms \cite{watson2018programmable, zwerver2022qubits, takeda2016fault, philips2022universal, undseth2023hotter}. Operating the devices at higher temperatures has been found to help mitigate this effect, suggesting that the resonance shift is caused by a heating of the device from the AC driving field \cite{undseth2023hotter}. Although the microscopic physics underpinning PIRS is still unknown, it has been suggested that it may be related to two-level systems in the qubit's environment \cite{choi2024interacting}. \\

Fig. \ref{fig:pulse_induced} shows the implications of PIRS on the fidelity of the nuclear geometric CZ gate. The pulse sequence schematic shown in Fig. \ref{fig:pulse_induced} \textbf{A.} shows the pulses carried out in this experiment. The nuclei were first initialized in the state $\ket{\Downarrow_1 \Downarrow_2}$. A $\frac{\pi}{2}$ was then applied to n1, such that the following nuclear superposition state was prepared: $\frac{1}{\sqrt{2}}(\ket{\Downarrow_1 \Downarrow_2} + \ket{\Uparrow_1 \Downarrow_2})$. A resonant ESR pulse was then applied to electron 2, conditional on the nuclear state $\ket{\Uparrow_1 \Downarrow_2}$, imparting a phase to this component of the nuclear superposition. The duration of this ESR pulse was swept between a range of $0-12\pi$, where $\pi$ refers to the $\pi$ time of e2, resulting in a varying geometric phase imparted on the nucleus. Finally, a $\frac{\pi}{2}$ pulse was again applied to n1, projecting the information along the X direction of the Bloch sphere to the Z-axis. For the case of an ESR duration of 0, no phase is imparted onto the nuclei and thus the nucleus is flipped from the $\ket{\Downarrow_1}$ to the $\ket{\Uparrow_1}$ state by the two $\frac{\pi}{2}$ pulses, resulting in $P_{\rm flip}=1$. Conversely, for an ESR duration of $2\pi n$, where $n$ is an integer, a phase of $\pi$ is imparted onto n1 and thus the nucleus will return to the $\Downarrow_1$ state upon the implementation of the final $\frac{\pi}{2}$ pulse, resulting in $P_{\rm flip}=0$. \\

Fig. \ref{fig:pulse_induced} \textbf{A}, shows the experimental implementation of this pulse sequence. As a control, the pulse sequence was also performed without the rotation of the electron (green points). The behaviour of the nucleus does not follow the expected trend of oscillating with a period of $2\pi$, indicating that the nuclear geometric CZ gate is not being correctly implemented. We believe the reason for this to be the PIRS observed in this device, whereby the electron resonance frequency shifts after the application of the first NMR pulse, before slowly drifting back to its original frequency during the ESR rotation. Fig. \ref{fig:pulse_induced} \textbf{B}, shows the implementation of the same pulse sequence but this time, preceded by a long, off-resonant NMR `saturation pulse'. The saturation pulse saturates the PIRS effect. By calibrating the ESR drive, such that it is on-resonance with this saturated frequency, we ensure that we start the ESR rotation on resonance. In this case the nuclear flip probability falls to 0 for a rotation duration of $2\pi$ on the electron, as expected, indicating the successful implementation of the nuclear geometric CZ gate. However, as the drive duration of the electron increases, the electron resonance frequency continues to return to its unsaturated frequency, resulting in the deviation from the expected behaviour beyond a rotation of $2\pi$.\\

To prove that the behaviour observed in this experiment does indeed originate from the effect of PIRS, we performed a simulation of this pulse sequence, whereby we exponentially swept the detuning of the microwave pulse applied to the electron from a detuning of 0 to 120 kHz, over the course of the maximum electron drive duration of $12\pi$. Although the exact frequency drift behaviour is difficult to fully re-create in simulations, the broad behaviour of the experiment seems to be captured by this model, indicating that PIRS is indeed responsible for the reduction in CZ fidelity as the ESR duration is increased. \\

To combat this experimentally, a saturation pulse was always applied before the implementation of the nuclear geometric CZ gate. 

\subsection*{Flip probability vs up proportion}

\begin{figure}[ht]
    \centering
    \includegraphics[width=0.8\textwidth] {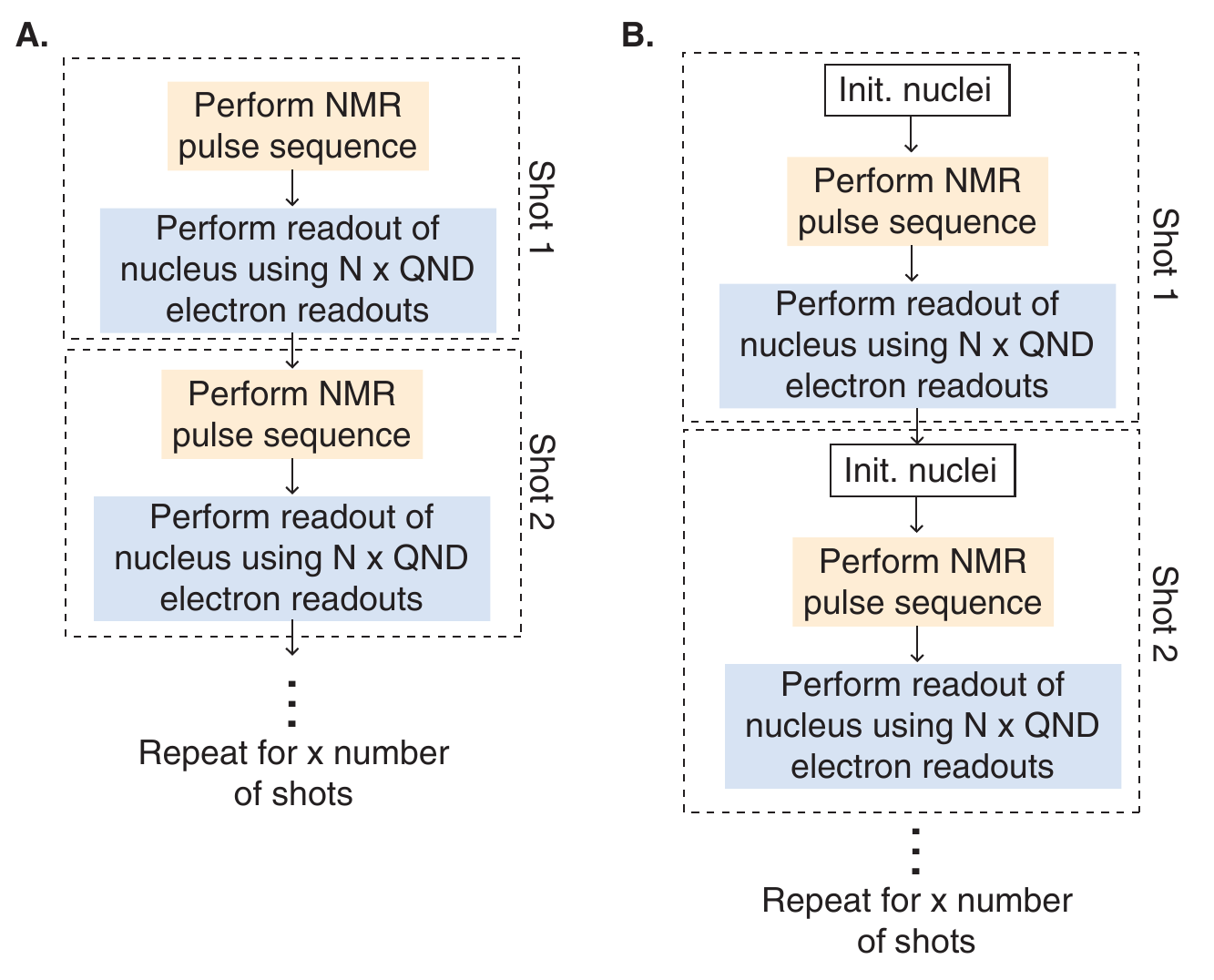}
    \caption[Flip probability vs up proportion processes]{\textbf{Flip probability vs up proportion processes.} \textbf{A.} Process flow of calculating the nuclear flip probability. \textbf{B.} Process flow of calculating the nuclear up proportion.}
    \label{fig:flip_prob}
\end{figure}

Throughout this work the nuclei were measured either by looking at nuclear flip probability, $P_{\rm flip}$, or nuclear up proportion (or state probability in the case of two-qubit readout), $P_{\Uparrow}$. Fig. \ref{fig:flip_prob} \textbf{A} shows the process flow behind measuring $P_{\rm flip}$. This process involves performing an NMR pulse sequence, followed by a QND readout of the nucleus via the electron, which together constitute a single `shot' of the measurement. The NMR pulse sequence, followed by QND readout of the nucleus is then repeated multiple times, thus resulting in multiple measurement 'shots'. $P_{\rm flip}$ is then given by $\frac{N_F}{(N-1)}$ where $N_F$ represents the number of times the measured nuclear state flipped between consecutive shots (e.g. from an outcome of 0 to 1 or from 1 to 0 between consecutive shots). $N$ represents the number of shots taken, with $N-1$ thus representing the maximum number of possible flips given this shot number. Flip probability is a useful measurement tool for the nucleus, as it does not require nuclear initialization, resulting in $P_{\rm flip}$ typically giving a higher readout contrast, as a consequence of the reduced state preparation error. \\

$P_{\Uparrow}$ is instead calculated by performing the same procedure as nuclear flip probability (see Fig. \ref{fig:flip_prob} \textbf{B}) however, in this case the nuclei are initialized before every shot, which consists of an NMR pulse sequence followed by QND readout using the electron. These shots are repeated many times and $P_{\Uparrow}$ is calculated with $\frac{N_{\Uparrow}}{N}$ where $N_{\Uparrow}$ are the number of shots for which the nucleus was measured to be in the $\ket{\Uparrow}$ state and $N$ represents the total number of shots. The measurement of $P_{\Uparrow}$ over $P_{\rm flip}$ is necessary in some measurements (for example when performing nuclear two-qubit measurements) however, it has the disadvantage of requiring nuclear initialization before every shot and therefore typically possesses a lower readout contrast than $P_{\rm flip}$. \\

Throughout this manuscript we have highlighted whether $P_{\rm flip}$ or $P_{\Uparrow}$ is measured for each experiment performed on the nuclei.

\subsection*{Nuclear one-qubit gate set tomography (GST)}

\begin{figure}[ht]
    \centering
    \includegraphics[width=1\textwidth] {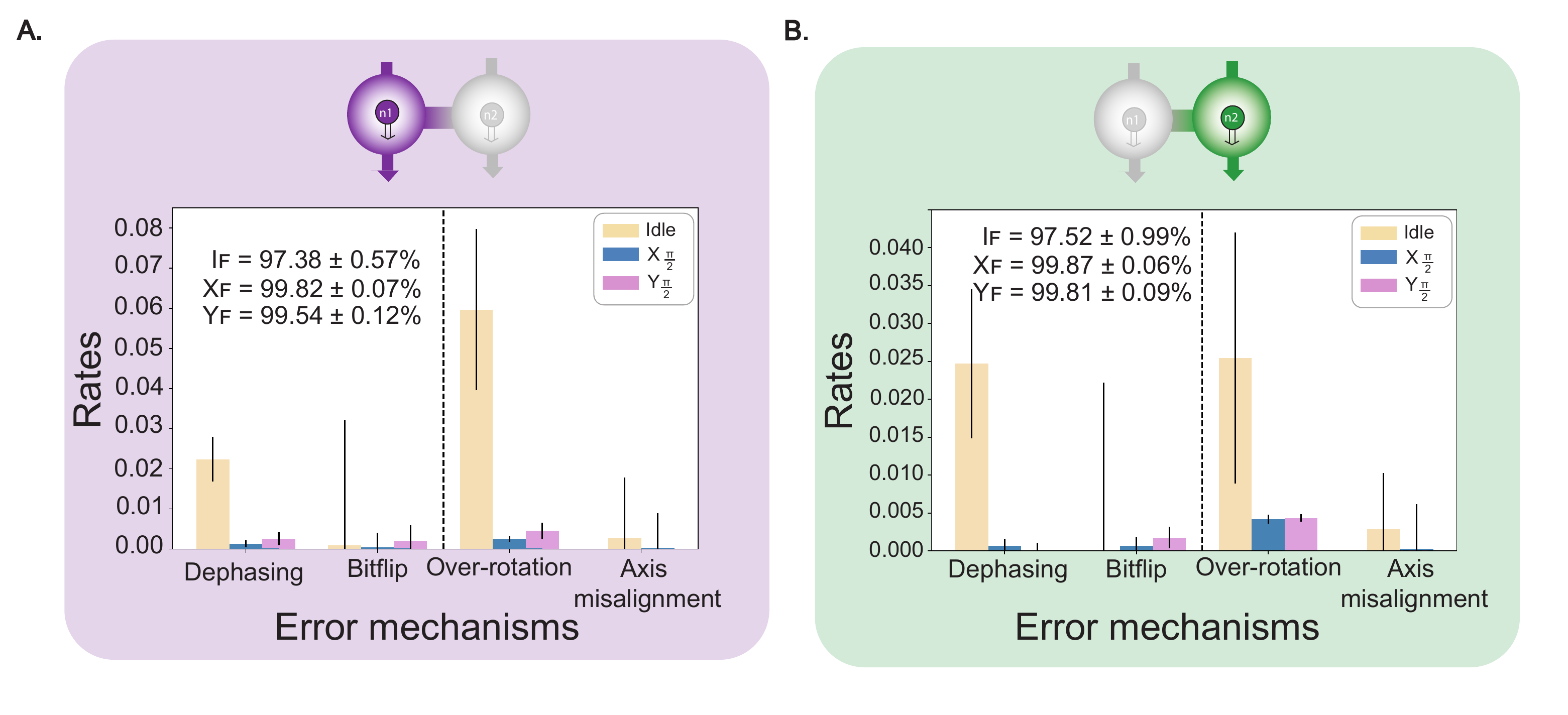}
    \caption[Nuclear one-qubit GST.]{\textbf{Nuclear one-qubit GST.} \textbf{A.} Characterisation of the single-qubit gates on n1, using gate set tomography (GST). The error generators were combined in order to estimate the error contribution from the four physical error mechanisms. The error mechanisms representing stochastic error processes are to the left of the black vertical dashed line, while the coherent error processes are to the right of the black dashed line. The inset shows the fidelities for the idle gate ($I_F$), $X_{\frac{\pi}{2}}$ gate ($X_F$) and $Y_{\frac{\pi}{2}}$ gate ($Y_F$). \textbf{B.} Characterisation of the single-qubit gates on n2, using GST.}
    \label{fig:one_qubit_GST}
\end{figure}

We characterised the single-qubit nuclear gates by performing gate set tomography (GST). A gateset consisting of an idle gate, $X_{\frac{\pi}{2}}$, and $Y_{\frac{\pi}{2}}$ gate were tested for both n1 and n2, up to a maximum circuit depth of $L = 512$. Fig. \ref{fig:one_qubit_GST} shows the error budgets for each gate broken down into four physical error mechanisms: dephasing, bitflip, over-rotation and axis-misalignment errors. These error rates were calculating using combinations of error generators provided by the GST analysis \cite{stemp2024tomography}. The insets of Fig. \ref{fig:one_qubit_GST} \textbf{A} and \textbf{B} show the estimated fidelities for each single-qubit gate, which are all above 99$\%$ for both nuclei, except for the idle gate. The overall model violation was 14.1 for n1 and 13.0 for n2, indicating a high degree of Markovianity \cite{nielsen2021gate}.

\subsection*{Phase map simulation}

\begin{figure}[ht]
    \centering
    \includegraphics[width=1\textwidth] {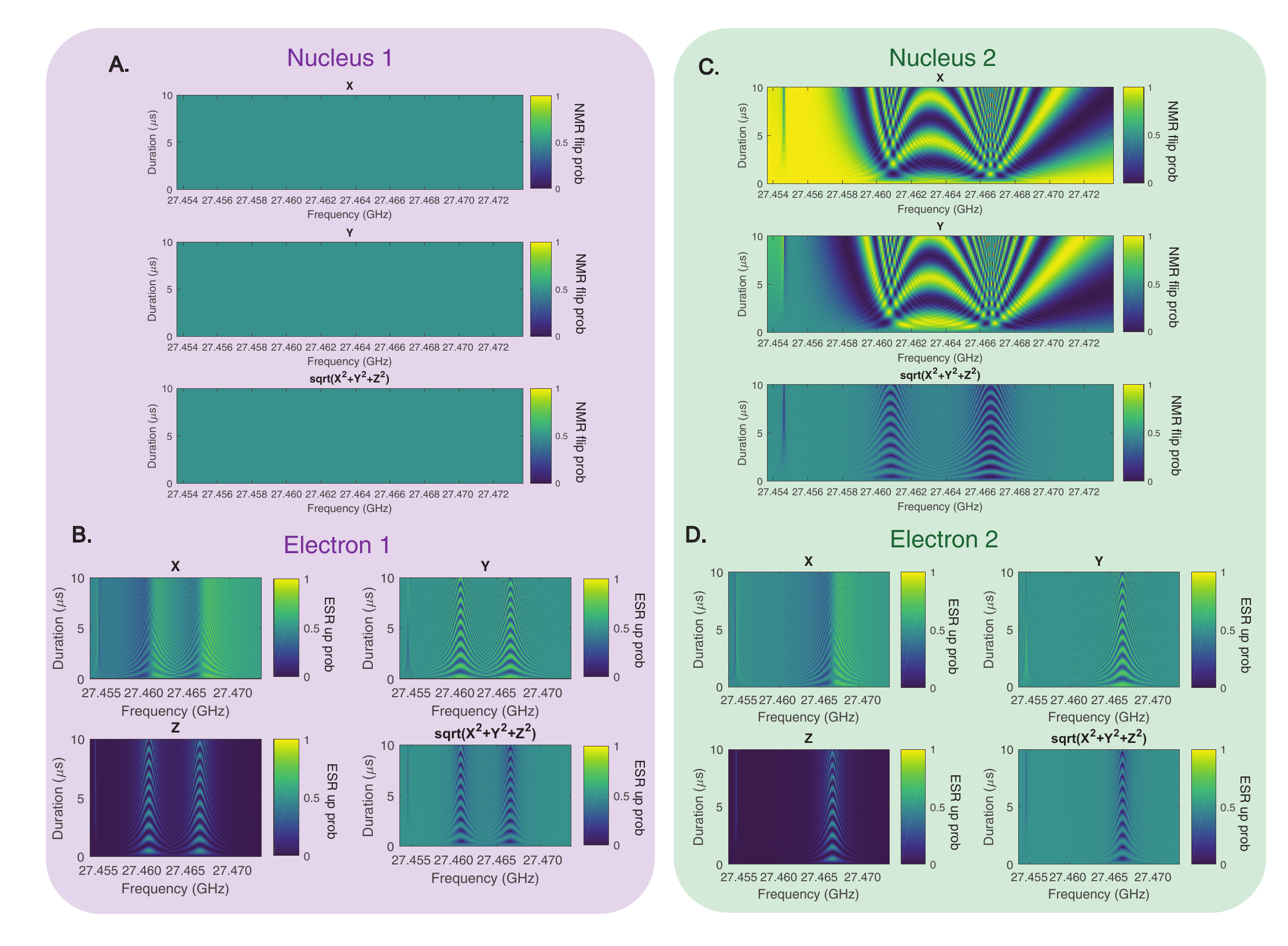}
    \caption[Simulation of the Control-Z gate.]{\textbf{Simulation of the Control-Z gate.} Simulation of the rotation of electron 1 with a varying duration and frequency, when the nuclei are in the superposition state $\frac{1}{\sqrt{2}}(\ket{\Downarrow_1 \Downarrow_1} + \ket{\Downarrow_1 \Uparrow_2})$. \textbf{A.} Simulation results of the up proportion along the X, Y and $\sqrt{X^{2} + Y^{2} + Z^{2}}$ direction for nucleus 1. \textbf{B.} Simulation results of the up proportion along the X, Y and $\sqrt{X^{2} + Y^{2} + Z^{2}}$ direction for nucleus 2. \textbf{C.} Simulation results of the up proportion along the X, Y and $\sqrt{X^{2} + Y^{2} + Z^{2}}$ direction for electron 1. \textbf{D.} Simulation results of the up proportion along the X, Y and $\sqrt{X^{2} + Y^{2} + Z^{2}}$ direction for electron 2. Note that the simulation parameters used to generate this figure differed slightly from those used in Fig.2 of the main text.}
    \label{fig:CZ_full_sim}
\end{figure}

Extended results of the simulation shown in Fig. 2 of the main text. The pulse sequence being simulated was the following:
\begin{enumerate}
    \item Initialize the nuclei in the superposition state $\frac{1}{\sqrt{2}}(\ket{\Downarrow_1 \Downarrow_1} + \ket{\Downarrow_1 \Uparrow_2})$.
    \item Apply an AC microwave field, where the duration of this microwave field is swept between 0-10 $\mu$s and the frequency swept $\pm$ 10 MHz about the centre frequency between the electron resonances for e1 conditional on the nuclear state $\ket{\Downarrow_1 \Uparrow_2}$ and $ \ket{\Downarrow_1 \Downarrow_2}$. This thus imparts a varying geometric phase to n2, which can be measured by calculating the expectation value of n2 along the X axis of the Bloch sphere. 
\end{enumerate}

The up proportion was calculated along the X, Y and $\sqrt{X^{2} + Y^{2} + Z^{2}}$ direction. The latter quantity is a good indicator of when the spins become entangled, as this quantity will tend towards 0 as the spins approach the center of the Bloch sphere when they become entangled. \\

Fig. \ref{fig:CZ_full_sim} \textbf{A} shows the results of the simulation for n1. This nucleus remains in the spin $\ket{\Downarrow_1}$ state throughout the pulse sequence and thus an up proportion of 0.5 is measured along the X, Y and $\sqrt{X^{2} + Y^{2} + Z^{2}}$ direction throughout.\\

Fig. \ref{fig:CZ_full_sim} \textbf{B} shows the results of the simulation for n2. As described extensively in the main text, $P_{\Uparrow}$ of this nucleus along the X and Y axis of the Bloch sphere depend on the phase imparted by the electron rotation. The value of $\sqrt{X^{2} + Y^{2} + Z^{2}}$ can be seen to go to 0 for this nucleus at the resonance frequencies of e1, conditional on the nuclear states $\ket{\Downarrow_1 \Uparrow_2}$ and $ \ket{\Downarrow_1 \Downarrow_2}$. This occurs every time the electron undergoes a rotation of $\pi$ at these resonances, indicating that n2 becomes entangled with the electrons at these points.\\

Fig. \ref{fig:CZ_full_sim} \textbf{C} shows the results of the simulation for e1. This electron undergoes rotations at the Bloch sphere when the rotation frequency approaches its resonance frequencies. The value of $\sqrt{X^{2} + Y^{2} + Z^{2}}$ is seen to go to 0 at these resonances whenever this electron undergoes a $\pi$ rotation, indicating that the electron has become entangled with n2.\\

Fig. \ref{fig:CZ_full_sim} \textbf{D} shows the results of the simulation for e2. When the nuclei are in a parallel spin-orientation then the resonance frequencies for e1 and e2 become degenerate. Thus, as the AC driving frequency approaches the frequency conditional on the nuclear state $\ket{\Downarrow_1 \Downarrow_2}$, both e1 and e2 are driven simultaneously. Consequently, a rotation of e2 is only observed close to this resonance frequency. When e2 undergoes a rotation of $\pi$ it also becomes entangled with n2, as indicated by $\sqrt{X^{2} + Y^{2} + Z^{2}}$ going to 0.

\subsection*{Bell state fidelity and concurrence}

Bell state tomography involves preparing a given Bell state, before reconstructing the density matrix of the prepared state, using state tomography, and comparing the resulting matrix to the expected density matrix.\\

Any qubit density matrix can be written as a superposition of the Pauli matrices, $\hat{\sigma}_{I} = \begin{pmatrix}
    1 & 0\\
    0 & 1
\end{pmatrix}, \hat{\sigma}_{x} = \begin{pmatrix}
    0 & 1\\
    1 & 0
\end{pmatrix}, \hat{\sigma}_{y} =  \begin{pmatrix}
    0 & -j\\
    j & 0
\end{pmatrix}, \hat{\sigma}_{z} = \begin{pmatrix}
    1 & 0\\
    0 & -1
\end{pmatrix}$ using the following: 

\begin{equation}
\label{density_from_stokes}
    \hat{\rho} = \frac{1}{2}I + \frac{1}{2}\sum_{i=1}^{i = 4^{n}-1}S_{i}\hat{\sigma}_{i},
\end{equation}

where $n$ represents the number of qubits in the system, $\sigma_{i} = \{ \hat{\sigma}_{I}, \hat{\sigma}_{x}, \hat{\sigma}_{y},\hat{\sigma}_{z} \}$ and $S_{i}$ are the Stokes parameters, which together form the Stokes vector, $\vec{S} = (S_{1}, S_{2}, S_{3},...S_{4^{n}-1})$ \cite{altepeter2005photonic, fano1954stokes}. For the case of a pure state, the magnitude of the Stokes vector is $|\vec{S}| = 1$. For a mixed state however, the state no longer lies on the surface of the Bloch sphere and instead shrinks towards the centre of the Bloch sphere, resulting in $|\vec{S}| < 1$.\\

The Stokes parameters can be calculated from the the density matrix $\hat{\rho}$ with the following formula

\begin{equation}
    S_{i} = \text{Tr}[\hat{\sigma}_{i}\hat{\rho}].
\end{equation}

In order to experimentally determine the Stokes parameters for a given system and hence reconstruct the density matrix we must therefore perform a set of measurements on the qubits in the three bases: $x,y$ and $z$. We can define $P_{\ket{\downarrow_{i}}}$($P_{\ket{\uparrow_{i}}}$) as the probability of obtaining the result $\ket{\downarrow_{i}}(\ket{\uparrow_{i}})$ upon measuring the qubit in the basis $\{ \ket{\downarrow_{i}},\ket{\uparrow_{i}} \}$. Using this notation, the Stokes parameters for the example of a single qubit can be calculated as the following

\begin{align*}
S_{0} = S_{ii} &= (P_{\ket{\downarrow_{z}}} + P_{\ket{\uparrow_{z}}}) \otimes (P_{\ket{\downarrow_{z}}} + P_{\ket{\uparrow_{z}}}) = P_{\ket{\downarrow_{z}\downarrow_{z}}} + P_{\ket{\downarrow_{z}\uparrow_{z}}} + P_{\ket{\uparrow_{z} \downarrow_{z}}} + P_{\ket{\uparrow_{z} \uparrow_{z}}},\\
S_{1} = S_{iz} &= (P_{\ket{\downarrow_{z}}} + P_{\ket{\uparrow_{z}}}) \otimes (P_{\ket{\downarrow_{z}}} - P_{\ket{\uparrow_{z}}}) = P_{\ket{\downarrow_{z}\downarrow_{z}}} - P_{\ket{\downarrow_{z}\uparrow_{z}}} + P_{\ket{\uparrow_{z} \downarrow_{z}}} - P_{\ket{\uparrow_{z} \uparrow_{z}}},\\
S_{2} = S_{ix} &= (P_{\ket{\downarrow_{z}}} + P_{\ket{\uparrow_{z}}}) \otimes (P_{\ket{\downarrow_{x}}} - P_{\ket{\uparrow_{x}}}) = P_{\ket{\downarrow_{z}\downarrow_{x}}} - P_{\ket{\downarrow_{z}\uparrow_{x}}} + P_{\ket{\uparrow_{z} \downarrow_{x}}} - P_{\ket{\uparrow_{z} \uparrow_{x}}},\\
S_{3} = S_{iy} &= (P_{\ket{\downarrow_{z}}} + P_{\ket{\uparrow_{z}}}) \otimes (P_{\ket{\downarrow_{y}}} - P_{\ket{\uparrow_{y}}}) = P_{\ket{\downarrow_{z}\downarrow_{y}}} - P_{\ket{\downarrow_{z}\uparrow_{y}}} + P_{\ket{\uparrow_{z} \downarrow_{y}}} - P_{\ket{\uparrow_{z} \uparrow_{y}}},\\
\\
S_{4} = S_{xi} &= (P_{\ket{\downarrow_{x}}} - P_{\ket{\uparrow_{x}}}) \otimes (P_{\ket{\downarrow_{z}}} + P_{\ket{\uparrow_{z}}}) = P_{\ket{\downarrow_{x}\downarrow_{z}}} + P_{\ket{\downarrow_{x}\uparrow_{z}}} - P_{\ket{\uparrow_{x} \downarrow_{z}}} - P_{\ket{\uparrow_{x} \uparrow_{z}}},\\
S_{5} = S_{xz} &= (P_{\ket{\downarrow_{x}}} - P_{\ket{\uparrow_{x}}}) \otimes (P_{\ket{\downarrow_{z}}} - P_{\ket{\uparrow_{z}}}) = P_{\ket{\downarrow_{x}\downarrow_{z}}} - P_{\ket{\downarrow_{x}\uparrow_{z}}} - P_{\ket{\uparrow_{x} \downarrow_{z}}} + P_{\ket{\uparrow_{x} \uparrow_{z}}},\\
S_{6} = S_{xx} &= (P_{\ket{\downarrow_{x}}} - P_{\ket{\uparrow_{x}}}) \otimes (P_{\ket{\downarrow_{x}}} - P_{\ket{\uparrow_{x}}}) = P_{\ket{\downarrow_{x}\downarrow_{x}}} - P_{\ket{\downarrow_{x}\uparrow_{x}}} - P_{\ket{\uparrow_{x} \downarrow_{x}}} + P_{\ket{\uparrow_{x} \uparrow_{x}}},\\
S_{7} = S_{xy} &= (P_{\ket{\downarrow_{x}}} - P_{\ket{\uparrow_{x}}}) \otimes (P_{\ket{\downarrow_{y}}} - P_{\ket{\uparrow_{y}}}) = P_{\ket{\downarrow_{x}\downarrow_{y}}} - P_{\ket{\downarrow_{x}\uparrow_{y}}} - P_{\ket{\uparrow_{x} \downarrow_{y}}} + P_{\ket{\uparrow_{x} \uparrow_{y}}},\\
\\
S_{8} = S_{yi} &= (P_{\ket{\downarrow_{y}}} - P_{\ket{\uparrow_{y}}}) \otimes (P_{\ket{\downarrow_{z}}} + P_{\ket{\uparrow_{z}}}) = P_{\ket{\downarrow_{y}\downarrow_{z}}} + P_{\ket{\downarrow_{y}\uparrow_{z}}} - P_{\ket{\uparrow_{y} \downarrow_{z}}} - P_{\ket{\uparrow_{y} \uparrow_{z}}},\\
S_{9} = S_{yz} &= (P_{\ket{\downarrow_{y}}} - P_{\ket{\uparrow_{y}}}) \otimes (P_{\ket{\downarrow_{z}}} - P_{\ket{\uparrow_{z}}}) = P_{\ket{\downarrow_{y}\downarrow_{z}}} - P_{\ket{\downarrow_{y}\uparrow_{z}}} - P_{\ket{\uparrow_{y} \downarrow_{z}}} + P_{\ket{\uparrow_{y} \uparrow_{z}}},\\
S_{10} = S_{yx} &= (P_{\ket{\downarrow_{y}}} - P_{\ket{\uparrow_{y}}}) \otimes (P_{\ket{\downarrow_{x}}} - P_{\ket{\uparrow_{x}}}) = P_{\ket{\downarrow_{y}\downarrow_{x}}} - P_{\ket{\downarrow_{y}\uparrow_{x}}} - P_{\ket{\uparrow_{y} \downarrow_{x}}} + P_{\ket{\uparrow_{y} \uparrow_{x}}},\\
S_{11} = S_{yy} &= (P_{\ket{\downarrow_{y}}} - P_{\ket{\uparrow_{y}}}) \otimes (P_{\ket{\downarrow_{y}}} - P_{\ket{\uparrow_{y}}}) = P_{\ket{\downarrow_{y}\downarrow_{y}}} - P_{\ket{\downarrow_{y}\uparrow_{y}}} - P_{\ket{\uparrow_{y} \downarrow_{y}}} + P_{\ket{\uparrow_{y} \uparrow_{y}}},\\
\\
S_{12} = S_{zi} &= (P_{\ket{\downarrow_{z}}} - P_{\ket{\uparrow_{z}}}) \otimes (P_{\ket{\downarrow_{z}}} + P_{\ket{\uparrow_{z}}}) = P_{\ket{\downarrow_{z}\downarrow_{z}}} + P_{\ket{\downarrow_{z}\uparrow_{z}}} - P_{\ket{\uparrow_{z} \downarrow_{z}}} - P_{\ket{\uparrow_{z} \uparrow_{z}}},\\
S_{13} = S_{zz} &= (P_{\ket{\downarrow_{z}}} - P_{\ket{\uparrow_{z}}}) \otimes (P_{\ket{\downarrow_{z}}} - P_{\ket{\uparrow_{z}}}) = P_{\ket{\downarrow_{z}\downarrow_{z}}} - P_{\ket{\downarrow_{z}\uparrow_{z}}} - P_{\ket{\uparrow_{z} \downarrow_{z}}} + P_{\ket{\uparrow_{z} \uparrow_{z}}},\\
S_{14} = S_{zx} &= (P_{\ket{\downarrow_{z}}} - P_{\ket{\uparrow_{z}}}) \otimes (P_{\ket{\downarrow_{x}}} - P_{\ket{\uparrow_{x}}}) = P_{\ket{\downarrow_{z}\downarrow_{x}}} - P_{\ket{\downarrow_{z}\uparrow_{x}}} - P_{\ket{\uparrow_{z} \downarrow_{x}}} + P_{\ket{\uparrow_{z} \uparrow_{x}}},\\
S_{15} = S_{zy} &= (P_{\ket{\downarrow_{z}}} - P_{\ket{\uparrow_{z}}}) \otimes (P_{\ket{\downarrow_{y}}} - P_{\ket{\uparrow_{y}}}) = P_{\ket{\downarrow_{z}\downarrow_{y}}} - P_{\ket{\downarrow_{z}\uparrow_{y}}} - P_{\ket{\uparrow_{z} \downarrow_{y}}} + P_{\ket{\uparrow_{z} \uparrow_{y}}}.\\
\end{align*}

Experimentally, in order to measure the Bell states along the different projection axis, we must re-prepare the Bell state for each projection axis measured. In the donor spin system, we are only able to directly measure along the Z-axis of the spin using spin dependent tunneling. In order to measure the projection along the X or Y axis we must therefore append a projection pulse to the end of the Bell state preparation pulse, to map the information along the X or Y axis onto the Z-axis of the Bloch sphere. For example, to measure the projection along the +X axis, we can apply a $\frac{\pi}{2}$ projection pulse about the -Y axis (i.e. with a phase of -90$^{0}$) in order to rotate the information along the +X axis, to the +Z axis of the Bloch sphere, where it can then be read out via spin-dependent tunnelling.\\

Upon measuring the required Stokes parameters, the density matrix, $\hat{\rho}$, of the state can be reconstructed using equation \ref{density_from_stokes}. For the example of a two-qubit system the density matrix is calculated from the measured Stokes parameters with the following

\begin{align*}
&II = \frac{S_{ii}}{4S_{ii}} \hat{\sigma_{I}} \otimes \hat{\sigma_{I}}, IZ  = \frac{S_{iz}}{4S_{ii}} \hat{\sigma_{I}} \otimes \hat{\sigma_{z}}, 
IX  = \frac{S_{ix}}{4S_{ii}} \hat{\sigma_{I}} \otimes \hat{\sigma_{x}},
 IY = \frac{S_{iy}}{4S_{ii}} \hat{\sigma_{I}} \otimes \hat{\sigma_{y}} \\
& XI = \frac{S_{xi}}{4S_{ii}} \hat{\sigma_{x}} \otimes \hat{\sigma_{I}},
 XZ = \frac{S_{xz}}{4S_{ii}} \hat{\sigma_{x}} \otimes \hat{\sigma_{z}},
 XX = \frac{S_{xx}}{4S_{ii}} \hat{\sigma_{x}} \otimes \hat{\sigma_{x}},
 XY = \frac{S_{xy}}{4S_{ii}} \hat{\sigma_{x}} \otimes \hat{\sigma_{y}} \\
& YI = \frac{S_{yi}}{4S_{ii}} \hat{\sigma_{y}} \otimes \hat{\sigma_{I}},
YZ  = \frac{S_{yz}}{4S_{ii}} \hat{\sigma_{y}} \otimes \hat{\sigma_{z}},
YX  = \frac{S_{yx}}{4S_{ii}} \hat{\sigma_{y}} \otimes \hat{\sigma_{x}},
YY  = \frac{S_{yy}}{4S_{ii}} \hat{\sigma_{y}} \otimes \hat{\sigma_{y}} \\
& ZI = \frac{S_{zi}}{4S_{ii}} \hat{\sigma_{z}} \otimes \hat{\sigma_{I}},
 ZX = \frac{S_{zx}}{4S_{ii}} \hat{\sigma_{z}} \otimes \hat{\sigma_{x}},
 ZZ = \frac{S_{zz}}{4S_{ii}} \hat{\sigma_{z}} \otimes \hat{\sigma_{z}},
ZY  = \frac{S_{zy}}{4S_{ii}} \hat{\sigma_{z}} \otimes \hat{\sigma_{y}}.
\end{align*}
\\

\begin{align}
    \hat{\rho} &= II + IZ + IX + IY \\
    & + XI +XZ + XX + XY \\
    & + YI + YZ + YX + YY \\
    &+ ZI + ZX + ZZ + ZY
\end{align}

Using this method, the following density matrix was reconstructed for the two-qubit nuclear Bell state. Note that as state preparation and measurement (SPAM) errors were not extracted in these measurements, the density matrix directly reconstructed from experimental data does not necessarily represent a physical density matrix. We therefore used an optimizer to obtain the closest physical density matrix to the reconstructed density matrix. We display this physical density matrix below.

\begin{equation}
  \hat{\rho} = \begin{pmatrix}
0.0798 + 0.0000i & 0.0918 - 0.0368i & 0.0708 - 0.0537i & 0.0267 - 0.0743i \\
0.0918 + 0.0368i & 0.3463 + 0.0000i & 0.3733 - 0.1142i & 0.1213 + 0.0105i \\
0.0708 + 0.0537i & 0.3733 + 0.1142i & 0.4503 + 0.0000i & 0.1090 + 0.0734i \\
0.0267 + 0.0743i & 0.1213 - 0.0105i & 0.1090 - 0.0734i & 0.1236 + 0.0000i
\end{pmatrix}
\end{equation}

We then calculated the fidelity of the Bell state, $F$, with the following:

\begin{equation}
    F = \langle \psi | \hat{\rho}| \psi \rangle =  76  ^{+5}_{-5}\% 
\end{equation}

where 

\begin{equation}
 \ket{\psi} = \frac{1}{\sqrt{2}}
    \begin{pmatrix}
        0 \\
        1\\
        1\\
        0
    \end{pmatrix},
\end{equation}

To calculate the value of concurrence for this prepared state we first obtain the following 

\begin{equation}
    \hat{R} = \sqrt{\sqrt{\hat{\rho}}\tilde{\rho}\sqrt{\hat{\rho}}}.
\end{equation}

where $\hat{\rho}$ is the measured density matrix and $\tilde{\rho}$ is the result of applying the spin flip operation to $\hat{\rho}$, given by

\begin{equation}
    \tilde{\rho} = (\hat{\sigma}_{y} \otimes \hat{\sigma}_{y})\hat{\rho}^*(\hat{\sigma}_{y} \otimes \hat{\sigma}_{y}),
\end{equation}

where $\hat{\rho}^*$ is the complex conjugate transpose of the measured density matrix. The value of concurrence is then given by 

\begin{equation}
    C= max\{0,\lambda_1 - \lambda_2 - \lambda_3 - \lambda_4\},
\end{equation}

where $\lambda_{i}$ are the square roots of the eigenvalues of $\hat{R}$ in decreasing order.

\subsection*{Error bar estimation}

\begin{figure}[ht]
    \centering
    \includegraphics[width=1\textwidth] {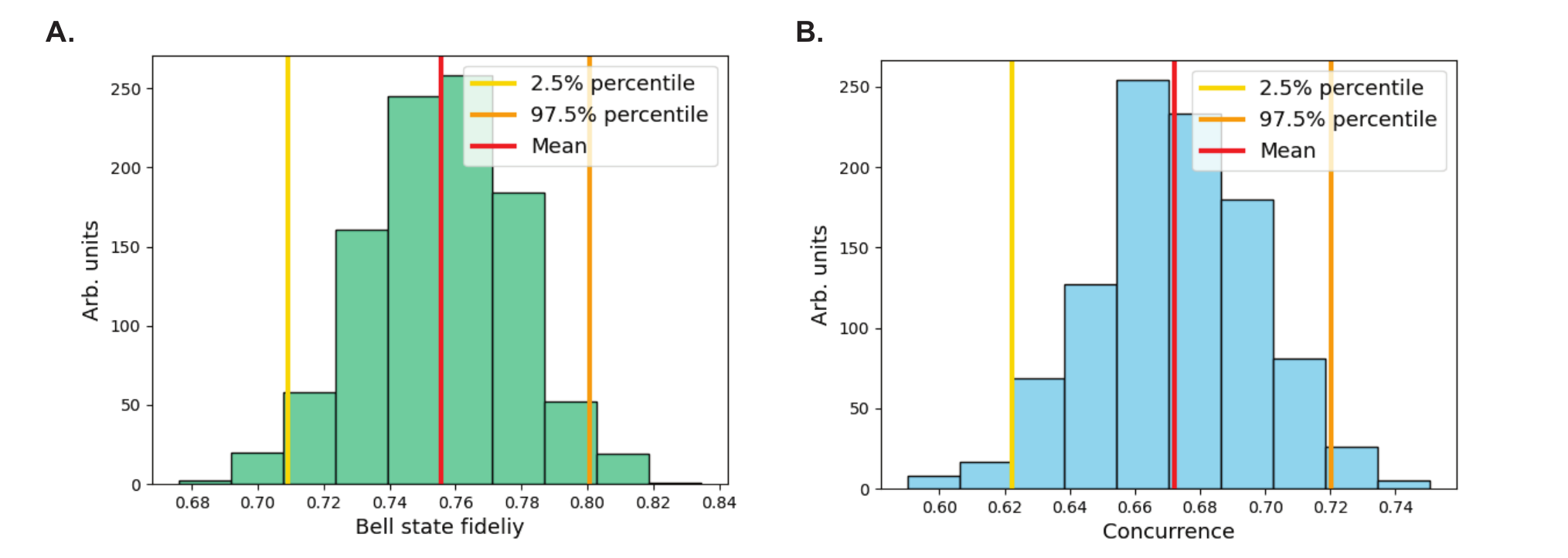}
    \caption[Bell state fidelity and concurrence error bars.]{\textbf{Bell state fidelity and concurrence error bars.} \textbf{A.} Histogram of Bell state fidelity calculated using non-parametric bootstrapping. The error bars were calculated using the difference between the values at the 2.5$\%$ and 97.5$\%$ percentile. \textbf{B.} Histogram of concurrence of the two-nuclei, calculated using non-parametric bootstrapping. The error bars were calculated using the difference between the values at the 2.5$\%$ and 97.5$\%$ percentile.}
    \label{fig:concurrence_bars}
\end{figure}

The error bars for the Bell state fidelity and concurrence values were calculated using a method of non-parametric bootstrapping. $N$ = 100 repetitions were taken of each Bell state tomography pulse sequence. The error bars were calculated by first dividing these 100 repetitions into 5 $\times$ 20 repetitions. 5 Bell state density matrices were then reconstructed for each of the groups of 20 repetitions. A non-parametric bootstrapping technique was then used to generate 1000 lists of 5 density matrices, sampled from the measured list. The mean of these lists were then plotted in a histogram (see Fig. \ref{fig:concurrence_bars} \textbf{A, B}) and the error bars were obtained by calculating the 2.5$\%$ and 97.5$\%$ percentile of the histograms.

\subsection*{Phase reversal tomography}
Another method to confirm entanglement between the two nuclei is through phase reversal tomography. Reversing the gates used to entangle the spins, while sweeping the phase of the pulses, allows for the reconstruction of the off-diagonal elements of the density matrix of the intended Bell state. One key advantage of this technique is that it allows for entanglement benchmarking in a system where the constituent components of the entanglement have vastly different coherence times. This case is true for the donor system, where the nuclear coherence time is approximately 2-orders of magnitude longer then that of the electron. By entangling the electron spin last and dis-entangling it first, the electron decoherence has limited impact on the fidelity of the benchmarking operations.
\\

\begin{figure}
    \centering
    \includegraphics[width=0.9\linewidth]{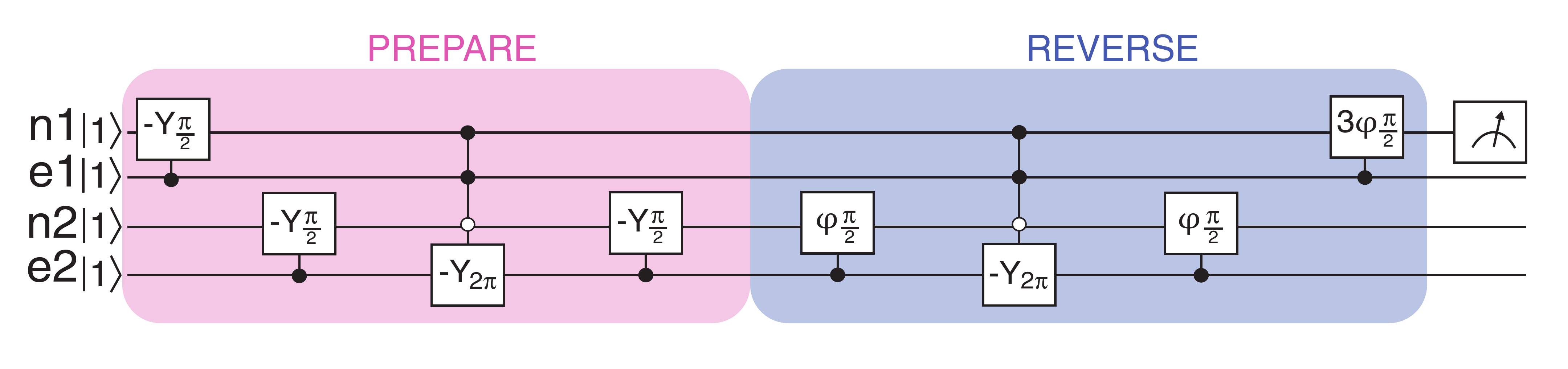}
    \caption{\textbf{Phase reversal tomography circuit.} Phase reversal tomography circuit including both nuclei ($\Downarrow$) and their respective electron underneath ($\downarrow$). In the prepare section all pulses are along the -Y axis, the $\phi$ in the reversal pulses is the axis of rotation that is swept from $0$ to $2\pi$. }
    \label{fig:phase-reversal-circuit}
\end{figure}

To show the principle behind this technique, we can consider only the behaviour of the nuclei and treat the nuclear CROT gate, consisting of two nuclear $\frac{\pi}{2}$ pulses and an electron $2\pi$ pulse, as a single gate. 
The decomposition of the CROT can be found in Fig. \ref{fig:phase-reversal-circuit}.\\

There are two gates necessary to generate a nuclear Bell state from an initial eigenstate. First we have a rotation on the first nucleus given by the unitary,
\begin{equation}
    R_1(\theta, \phi) = e^{-i\frac{\theta}{2}} {}^{ \begin{bmatrix}
        0 & e^{i\phi} \\
        e^{-i\phi} & 0 \\
    \end{bmatrix}}\otimes \hat{I}
    \label{eq:rot_matrix}
\end{equation}
the special case of interest when $\theta = \frac{\pi}{2}$ gives,
\begin{align}
    R_1(\pi/2, \phi) &=  \frac{1}{\sqrt{2}}\begin{bmatrix}
        1 & -ie^{i\phi} \\
        -ie^{-i\phi} & 1 \\
    \end{bmatrix} \otimes \hat{I}\\
    % &=\frac{1}{\sqrt{2}}\begin{bmatrix}
    %     1 & 0 & -ie^{i\phi} & 0\\
    %     0 & 1 &  0 & -ie^{i\phi}\\
    %     -ie^{-i\phi} & 0 & 1 & 0 \\
    %     0 & -ie^{-i\phi} & 0 & 1 \\
    % \end{bmatrix}.
    % \label{eq:Crot_matrix}
\end{align}

This results in the nuclei being initialized into a superposition state.\\

The second is the entangling gate, which in our case is performed on the second nucleus, n2, and is conditional on the state of the first nucleus, n1. This gate is given by,
\begin{equation}
        CR_2(\theta, \phi) =\ket{\Uparrow}\bra{\Uparrow}\otimes e^{-i\frac{\theta}{2}} {}^{ \begin{bmatrix}
        0 & e^{i\phi} \\
        e^{-i\phi} & 0 \\
    \end{bmatrix}} + \ket{\Downarrow}\bra{\Downarrow}\otimes \hat{I}.
    \label{eq:crot_matrix}
\end{equation}
where the special case of interest is $\theta = \pi$ which gives,
\begin{align}
        CR_2(\pi, \phi) &=\ket{\Uparrow}\bra{\Uparrow}\otimes -i \begin{bmatrix}
        0 & e^{i\phi} \\
        e^{-i\phi} & 0 \\
    \end{bmatrix} + \ket{\Downarrow}\bra{\Downarrow}\otimes \hat{I}\\
    % &=\begin{bmatrix}
    %     1 & 0 & 0  & 0\\
    %     0 & 1 &  0 & 0 \\
    %     0 & 0 & 0 & -ie^{i\phi} \\
    %     0 & 0 & -ie^{-i\phi} & 0 \\
    % \end{bmatrix}.
\end{align}
We can now generate a Bell state by applying both of these gates to our initial nuclear eigenstate,
\begin{equation}
\ket{\Psi_-} = CR_2(\pi, 0) R_1(\pi/2, 0) \ket{\Downarrow \Downarrow} = \hat{U} \ket{\Downarrow \Downarrow}
\end{equation}
Here we have defined $\hat{U}$ as the operator that creates the Bell state from the initial state $\ket{\Downarrow \Downarrow}$.
In phase reversal tomography, we reverse the sequence while sweeping the angles of rotation as,
\begin{align}
U_{R}\left(\phi_1, \phi_2\right) &= R_1\left(\pi/2, \phi_1\right)  CR_2\left(\pi, \phi_2\right)\\
\end{align}
Sweeping the phase of these reversal pulses we get
\begin{align}
P_{\Uparrow}\left(\phi_1, \phi_2\right) & =\frac{1}{2} + \frac{1}{2}\operatorname{tr}\left\{\hat{\sigma}_{z,1} \hat{U}_{R} \left(\phi_1, \phi_2\right) \ket{\Psi}\bra{\Psi} \hat{U}_{R}^{\dagger} \left(\phi_1, \phi_2\right) \right\} \\
& =\frac{1}{2} + \frac{1}{2}\cos(\phi_1 + \phi_2)
\end{align}

Deviations from the intended Bell state show up on the phase and amplitude of the oscillation of $P_\Uparrow$. 
To show why this is the case, we can apply the reversal circuit on the generalized off-diagonal elements of the Bell state density matrix alone. 
\begin{align}
P_{\Uparrow}\left(\phi_1, \phi_2\right) & =\frac{1}{2} + \frac{1}{2}\operatorname{tr}\left\{\hat{\sigma}_{z,1} \hat{U}_{R} \left(\phi_1, \phi_2\right) \begin{bmatrix}
        0 & 0 &  &  ae^{i\phi_B}\\
        0 & 0 &  0 & 0\\
        0 & 0 & 0 & 0 \\
         a^*e^{-i\phi_B} & 0 & 0 & 0 \\
    \end{bmatrix}
\hat{U}_{R}^{\dagger} \left(\phi_1, \phi_2\right) \right\} \\
& = \frac{a}{2}\cos(\phi_1 + \phi_2 - \phi_B)
\end{align}

The actual pulse sequence used is shown in Fig. \ref{fig:phase-reversal-circuit} and differs from the gates in Equations \ref{eq:rot_matrix} and \ref{eq:crot_matrix} in two ways. First, the nuclear rotations are conditional on their respective electrons. Ideally, both electrons remain in the down state throughout the circuit, making this conditionality negligible in principle. However, this assumption becomes relevant in the state preparation and measurement (SPAM) considerations discussed in the next section. Second, the CROT gate is not a native operation and is instead decomposed into a geometric CZ gate flanked by two $\frac{\pi}{2}$ rotations. Additionally, in the reversal pulses, the second phase is set to be three times the first, leading to four expected oscillations as the phase $\phi$ is swept from 0 to $2\pi$.

\begin{figure}
    \centering
    \includegraphics[width=\textwidth]{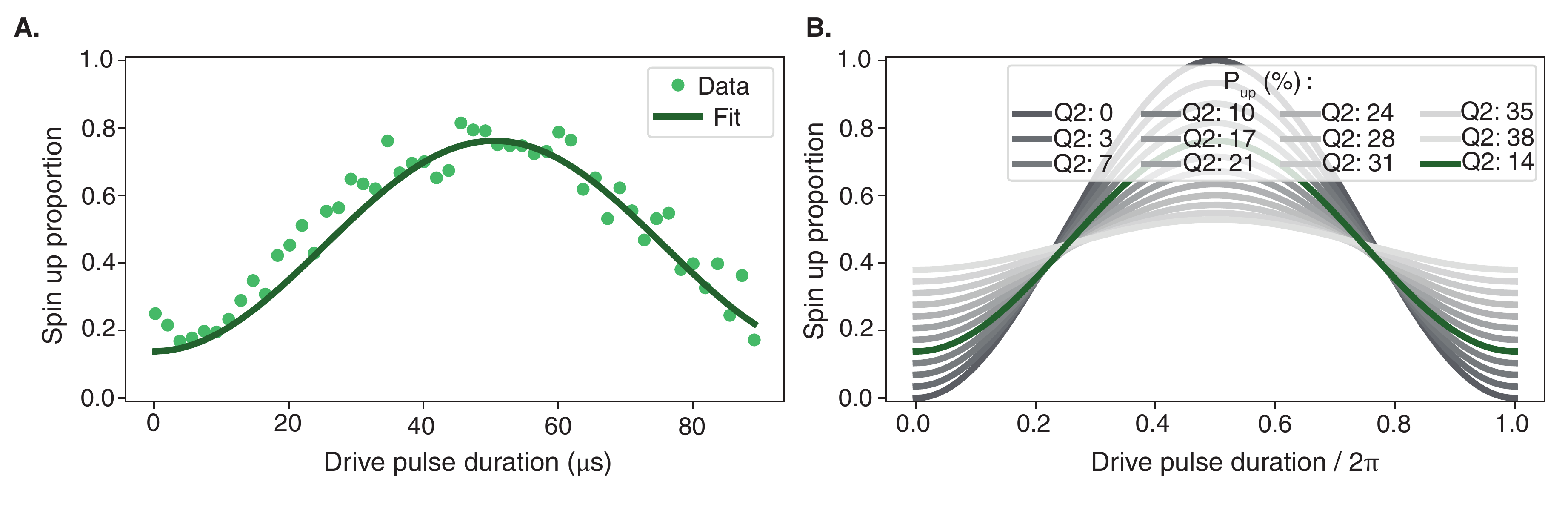}
    \caption{\textbf{Effect of erroneously loading an $\ket\uparrow$ electron on a nuclear Rabi oscillation.} \textbf{A.} Experimentally measured neutral nuclear Rabi oscillation that is fit to extract $P_{up}$. \textbf{B.} Simulations with initial state given by equation \ref{eq:full_initial_state}, where $P_{up}$ is swept.}
    \label{fig:spam-estimation-rabi}
\end{figure}

\begin{equation}
    P_{\Uparrow}(\phi) =\frac{1}{2} + \frac{1}{2}\operatorname{tr}\left\{\hat{\sigma}_{z,1} \hat{U}_{R} \left(\phi, 3\phi\right) \hat{U}
    \ket{\Downarrow \downarrow \Downarrow \downarrow}\bra{\Downarrow \downarrow \Downarrow \downarrow}
    \hat{U}^{\dagger} \hat{U}_{R}^{\dagger} \left(\phi, 3\phi\right) \right\} =\frac{1}{2} - \frac{1}{2}\cos(4\phi)
    \label{eq:phase-reversal}
\end{equation}

To get a better sense of the magnitude of the error in our phase reversal tomography that is associated with control error and the magnitude that is associated SPAM error, we will attempt to assess the effect of the SPAM error on the fidelity of the sequence.\\

Since the nuclear readout is quantum non-demolition (QND) we can perform repeated nuclear readout of the nuclear state for every shot.
We thus expect the measurement errors to be low and will focus on initialization errors. To do this, we will use a neutral Rabi oscillation performed on the nucleus to assess the electron initialization fidelity and then use simulations to determine the effect of this error on the phase reveresal circuit.\\

The nuclear initialization is realised through an ENDOR sequence, as explained in Ref \cite{stemp2024tomography}.
The sequence is based on the assumption that a newly loaded electron will tunnel onto the donor in the spin down state.
The ENDOR sequence then projects this down state onto one of the nuclei or the other electron of the exchange-coupled pair.
This means that if the read-out electron tunnels onto the donor in a spin up state, we erroneously project that spin up state onto the other spin during the initialization. 
If we thus have some finite probability $P_{up}$ of loading a spin up electron, the simplest model for initialization errors would be to assume that this initialization probability is the same for all the spins in our simulation.
Experimentally, we wish to start in a state consisting of all spins in the down state. We can thus express the true initialized state as the following mixed state.
\begin{equation}
    \ket{\Psi_0}\bra{\Psi_0} = P_{up} \ket{\Uparrow}\bra{\Uparrow} + (1 - P_{up}) \ket{\Downarrow}\bra{\Downarrow} 
    \label{eq:SPAM_state}
\end{equation}
\begin{figure}
    \centering
    \includegraphics[width=\textwidth]{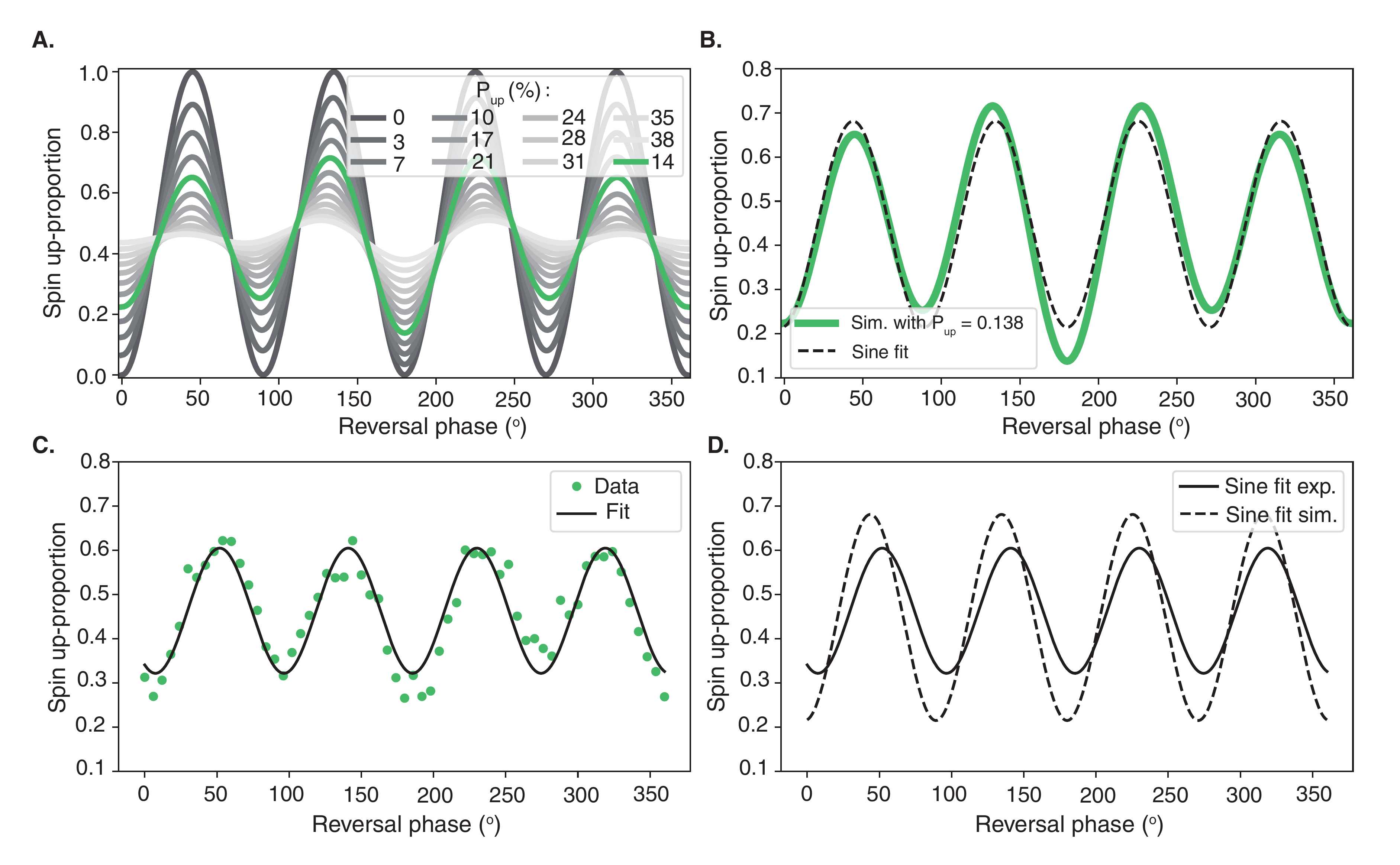}
    \caption{\textbf{A.} Simulation of circuit in Fig. \ref{fig:phase-reversal-circuit} with increasing erroneous spin up initialization probability ($P_{up}$). In green the $P_{up}$ value estimated from experiment (Fig. \ref{fig:spam-estimation-rabi}). \textbf{B.} Cosine fit of simulation. \textbf{C.} Cosine fit of experimental data. \textbf{D.} Comparison between fit of simulation and fit of data showcasing that the experimental fit has a phase offset of -0.638 rad and amplitude reduction of a factor 0.61 with respect to simulation.} 
    \label{fig:phase-reversal}
\end{figure}

The accidental electron spin up loading is a known issue in these devices and has been reported with a probability of 0.2 in very similar devices, \cite{johnson2022beating} in the absence of methods to improve this. This gives us a single parameter that captures the initialization errors. \\

For a two nucleus state this would be:
\begin{equation}
    \ket{\Psi_0}\bra{\Psi_0}  \otimes \ket{\Psi_0}\bra{\Psi_0}  = \begin{bmatrix} 
    (1-P_{up})^2 & 0 & 0 & 0 \\
    0 & P_{up}(1-P_{up}) & 0 & 0 \\ 
    0 & 0 & P_{up}(1-P_{up}) & 0 \\ 
    0 & 0 & 0 & P_{up}^2  \\ 
    \end{bmatrix}
\end{equation}

The loading of electrons into the spin-up state impacts more than just the initialization of the nuclear spins. The NMR frequencies are conditional on the state of the electron to which it is coupled via the hyperfine interaction. The NMR resonances driven in our sequences are conditional on the bound electron being in the down state. If the electron is in the up state then this results in the NMR drive being off-resonance.
Looking at a neutral Rabi oscillation we have the amplitude going down due to the nucleus starting in the opposite state with probability $P_{up}$ and, additionally, a reduction due to the NMR drive being off resonance, with a probability of $P_{up}$.

To model the full effect we need the full four spin system with equal initialization error probabilities
\begin{equation}
    D_0 = \ket{\Psi_0}\bra{\Psi_0} \otimes \ket{\Psi_0}\bra{\Psi_0} \otimes \ket{\Psi_0}\bra{\Psi_0} \otimes \ket{\Psi_0}\bra{\Psi_0} ,
    \label{eq:full_initial_state}
\end{equation}
with $\ket{\Psi_0}$ like in equation \ref{eq:SPAM_state}.

We assume that just after retuning the resonance frequencies of the NMR through a Ramsey experiment the only errors in a Rabi oscillation are given by the electron spin up loading probability $P_{up}$.
We can then simulate the Rabi with this finite probability and fit to the data to extract $P_{up}$.
Note that this is not visible in Rabi oscillations with the y-axis measured in nuclear flip probability (see Supplementary Materials Section `Flip probability vs up proportion'). In these cases the nuclear spins are not initialized and hence the flip-probability does not include nuclear initialization error. \\

In Fig. \ref{fig:spam-estimation-rabi} we fit the data taken from a Rabi oscillation on the nucleus to extract a $P_{up}$ value of 0.14. As derived in equation \ref{eq:phase-reversal} we expect a sine wave with four periods. However, because our initialization errors have multiple effects on the circuit this shape is not preserved. 
The expected effect of the initialization errors can be seen in simulation by taking the trace from equation \ref{eq:phase-reversal} and substituting in the intended initial state ($\ket{\Downarrow \downarrow \Downarrow \downarrow}\bra{\Downarrow \downarrow \Downarrow \downarrow}$) with the density matrix $D_0$ from equation \ref{eq:full_initial_state}.
The result for different values of spin up loading ($P_{up}$) are shown in \ref{fig:phase-reversal} \textbf{A}. The value of $P_{up}$ extracted from the Rabi experiment is highlighted in green.
Thee simulation only includes initialization errors and no other operational errors or dephasing. We can thus compare the experimental result to the simulation in similar spirit to SPAM extraction.
One simple way to compare these two curves is to fit a sine through both and compare the respective amplitude and phase.
In the idealised case we expect these two values to give us the off-diagonal density matrix elements of the Bell state.

In Fig. \ref{fig:phase-reversal} \textbf{D} the sine fits are compared.
Comparing the results of the simulation to the experimental data, we observe a phase offset of -0.638 rad and amplitude reduction of a factor 0.61. This is indicative of errors in the gates themselves.

\subsection*{Electron initialization}
\begin{figure}[ht]
    \centering
    \includegraphics[width=0.7\textwidth] {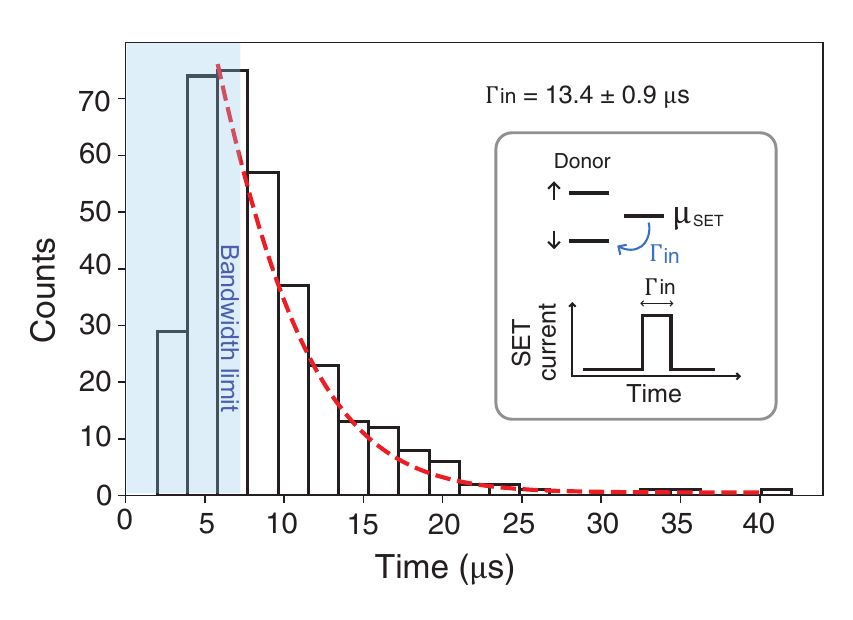}
    \caption[Tunnelling time calculation.]{\textbf{Tunnelling time calculation.} Histogram of the tunnelling time of the electron from the SET island to the donor taken from 500 measured SET current traces. The red line shows the exponential fit to this histogram, from which we extracted the tunnel time. Blue box shows the bandwidth limit set by the rise/fall time of the transimpedance amplifier. }
    \label{fig:tunneling}
\end{figure}

Although the nuclear Bell state fidelity was likely limited by poor electron initialization fidelity, this initialization fidelity could not be optimised using a Bayesian Maxwell's demon approach, as a result of the fast tunnel-in times of the electron from the SET to the donor. The Bayesian approach to initialization involves monitoring the SET current in real-time and triggering the start of a pulse sequence conditional on not observing a blip of current for a sufficient time. This time is chosen such that the confidence of the donor electron occupying the spin $\ket{\downarrow}$ state is above a desired threshold. The Bayesian initialization method has been shown to result in a 20x improvement in electron initialization fidelity and is now routinely incorporated into the electron initialization procedure in these devices \cite{johnson2022beating}. \\

In this device the electron tunnel time from the SET to the donor was measured to be 13.4 $\pm$ 0.9 $\mu$s (using the measurement shown in Fig. \ref{fig:tunneling}). This tunnelling time directly corresponds to the duration of high current through the SET. Thus, as the tunnelling time approaches the limit of the instrument bandwidth set by the rise and fall time of the transimpedance amplifier (7 $\mu$s in the present experiment), the probability increases of the SET current not rising above the threshold required to be counted as an electron spin $\ket{\uparrow}$. This has a detrimental effect on the effectiveness of the Bayesian update scheme, as missed current blips result in incorrect triggering, which can explain the observed lack of improved electron initialization fidelity using Bayesian update techniques in this device. This issue can therefore be mitigated in the future by using cryogenic transimpedance amplifiers with a higher bandwidth at an equivalent noise floor. Furthermore, it can also be addressed by using deterministic rather than timed ion-implantation, in order to drastically improve the precision with which the donors can be placed \cite{jakob2022deterministic}. This will ensure that the donor atoms reside within an optimal distance from the SET, such that the tunnelling time of the electron between the SET island and the donor is optimized.

\subsection*{Nuclear and electron coherence times}

\begin{figure}[ht]
    \centering
    \includegraphics[width=0.8\textwidth] {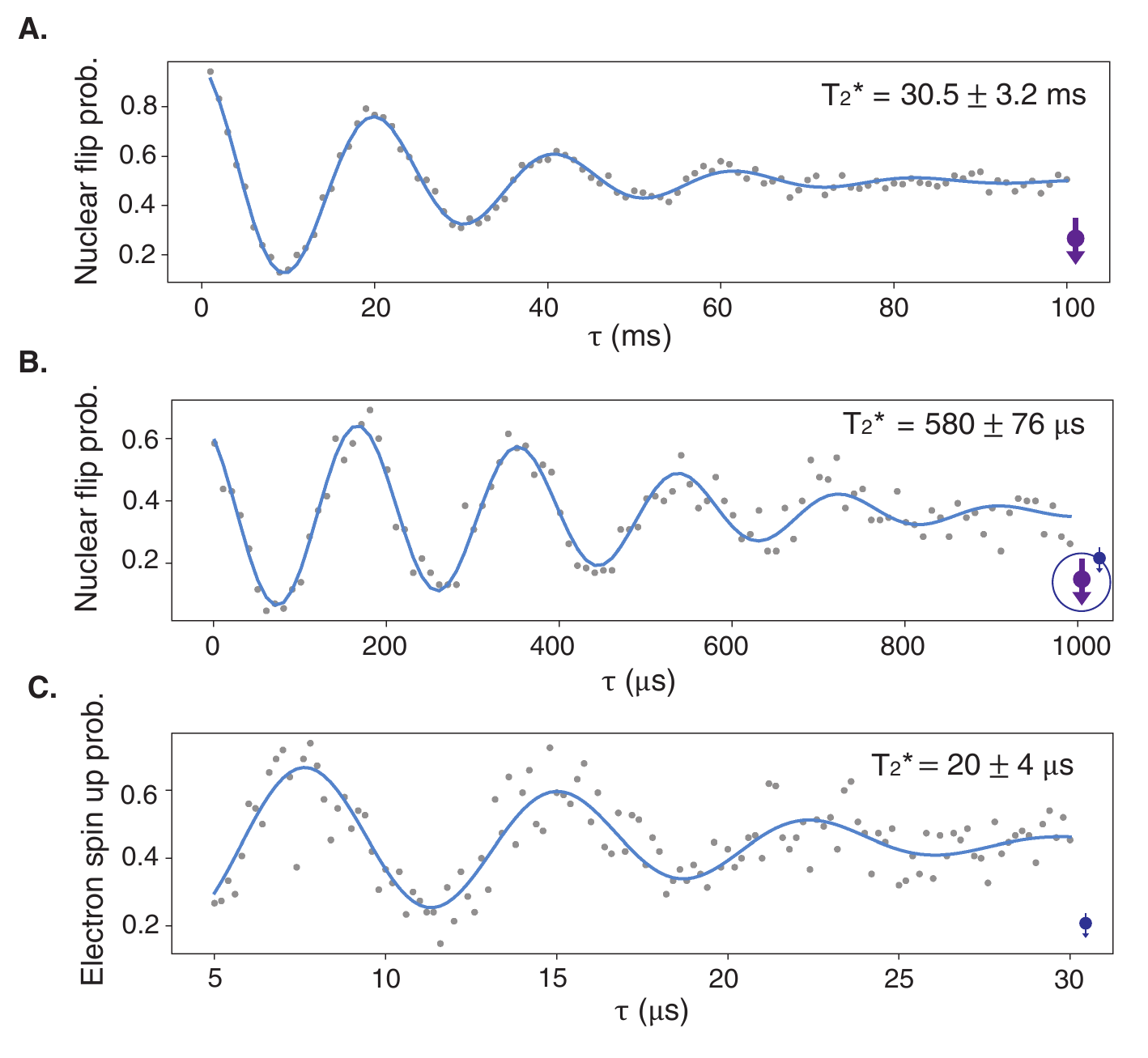}
    \caption[Coherence times.]{\textbf{Coherence times.} \textbf{A.} Ramsey experiment performed on the ionized nucleus 1. \textbf{B.} Ramsey experiment performed on the neutral nucleus 1. \textbf{C.} Ramsey experiment performed on electron 1. Error bars represent the 2$\sigma$ confidence interval extracted from the fit.}
    \label{fig:coherences}
\end{figure}

Fig. \ref{fig:coherences} shows Ramsey experiments performed on the ionized nucleus 1 (Fig. \ref{fig:coherences} \textbf{A.}), neutral nucleus 1 (Fig. \ref{fig:coherences} \textbf{B.}) and electron 1 (Fig. \ref{fig:coherences} \textbf{C.}). T$_{2}^{\ast}$ times of 30.5 $\pm$ 3.2 ms, 580 $\pm$ 76 $\mu$s and 20 $\pm$ 4 $\mu$s were extracted for the three spins respectively. The ionized nucleus is sensitive only to magnetic noise in its environment. The dominant source of magnetic noise that we expect in these devices is from residual $^{29}$Si nuclear spins in the vicinity of the donor. Although the silicon substrate was enriched with $^{28}$Si, there remains approximately 800 ppm of residual $^{29}$Si nuclear spins in the material. This magnetic noise can therefore be reduced through further isotopic enrichment of the silicon \cite{holmes2021isotopic} . The neutral nucleus and electron on the other hand are susceptible to both magnetic and electric noise in the environment, which can modulate both the  electron spin $g$ factor and the electron-nuclear hyperfine interaction. This electric noise arises from both noise associated with the electric gates used to control the donor atoms as well as charge-traps in the vicinity of the donor.

\subsection*{Inter-donor distance estimations}

\begin{figure}[ht]
    \centering
    \includegraphics[width=0.7\textwidth] {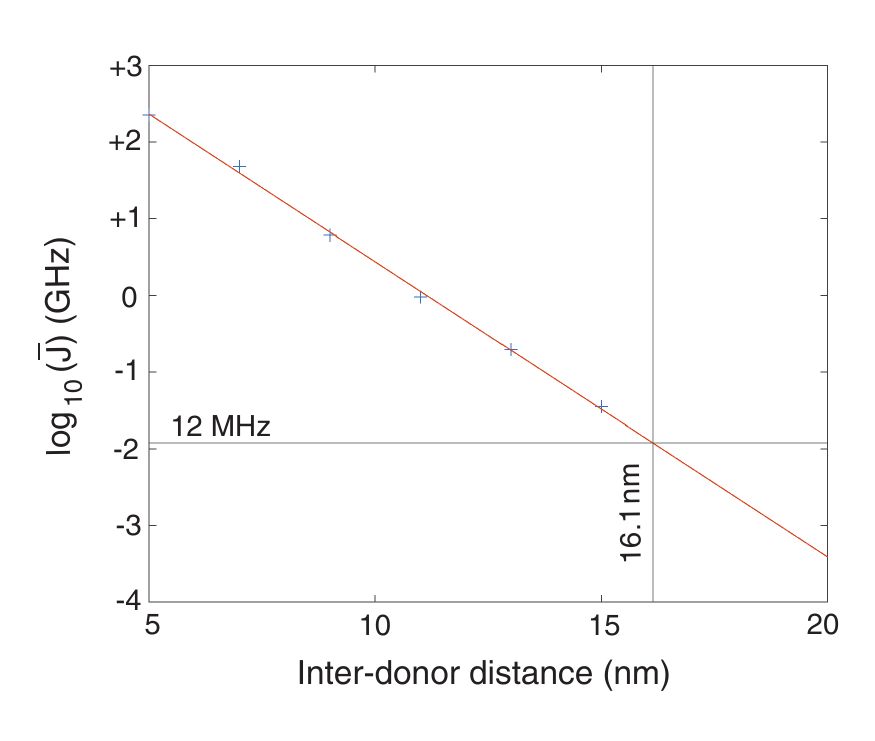}
    \caption[Inter-donor distance estimation.]{\textbf{Inter-donor distance estimation.} Data points represent simulated values of exchange-interaction strength from \cite{joecker2021full}, averaged over all possible donor orientation axis. The data is plotted on a logarithmic scale and fit with a linear fit. Vertical and horizontal lines represent the average inter-donor distance corresponding to 12 MHz of exchange interaction strength, which occurs at 16.1 nm.}
    \label{fig:donor_distance}
\end{figure}

The estimated inter-donor distance that corresponds to an exchange interaction strength of 12 MHz depends strongly upon which crystallographic directions the donors are separated. Without knowing along which axis the donors are separated, we can calculate an estimate of the inter-donor distance corresponding to 12 MHz of exchange, averaged over all possible donor orientations. To do this, we used full configuration interaction simulations from \cite{joecker2021full}, which evaluated the exchange-interaction strength along a sphere of possible donor-orientations, for different inter-donor distances. Averaging over all possible donor orientation axis, we plotted, on a logarithmic scale, the average exchange interaction for each donor separation distance, as shown in Fig. \ref{fig:donor_distance}. Performing a linear fit to this data we extracted the inter-donor distance at which the exchange-interaction, averaged across all donor orientations, corresponded to a value of 12 MHz. The inter-donor distance we extracted using this method was 16.1 nm.

\subsection*{Compatibility of this nuclear coupling scheme with deterministic ion implantation}
The two-qubit nuclear gate demonstrated in this work allows us to realize the scalability of this platform as a result of this coupling scheme being fundamentally compatible with deterministic ion implantation \cite{jakob2022deterministic}.\\

Since the precise value of the exchange interaction $J$ between the donor atoms is irrelevant for the nuclear coupling scheme outlined in this work, provided $J \ll A$ and larger than the resonance linewidth, this scheme is relatively insensitive to uncertainties in the precise location of the donors. The optimal range of donor distances for carrying out this two-qubit nuclear gate is between approximately 10-24 nm \cite{joecker2021full}. This is crucial, as it allows the coupling scheme to be robust against the uncertainty in donor placement intrinsic to the deterministic ion implantation process required to scale this platform.\\

The uncertainty in lateral donor placement during the ion implantation process has significantly improved in recent years, as a result of utilizing an atomic-force microscope (AFM) nanostencil tip, with an aperture through which the ion-beam can pass, to increase the localization of the donor implantation site. Using this procedure, the uncertainty in lateral donor placement depends on a number of factors, including: the lateral straggle of the donor as it is implanted in the lattice ($\approx$ 9 nm for a P ion implanted at a mean implantation depth of 20 nm and $\approx$ 5 nm for an Sb ion implanted at the same mean depth), the uncertainty in nanostencil aperture size ($\approx$ 5 nm for an aperture size of 10 nm) and the uncertainty in AFM positioning ($\approx$  0.5 nm). Combining these factors gives a rough estimate for the lateral uncertainty in donor placement of $\approx \pm$ 14.5 nm for a phosphorus donor (which can be further improved by implanting a molecular ion \cite{holmes2024improved}) and $\approx \pm$ 10.5 nm for an antimony donor \cite{jamieson2017deterministic}. This lateral uncertainty in donor placement, achievable with ion-implantation techniques, is therefore compatible with the range of viable distances for the optimal value of inter-donor distances for this scheme.

\end{document}